\documentclass[12pt]{article}

\usepackage[dvips]{graphicx}
\usepackage{amsmath}
\usepackage{amsfonts}
\usepackage{amssymb}

\newcommand{\be}{\begin{equation}}
\newcommand{\ee}{\end{equation}}
\newcommand{\bea}{\begin{eqnarray}}
\newcommand{\eea}{\end{eqnarray}}

\newcommand{\nn}{\nonumber}

\newcommand{\sign}{{\rm sign}}

\newcommand{\lp}{\left(}
\newcommand{\rp}{\right)}
\newcommand{\lb}{\left<}
\newcommand{\rb}{\right>}

\begin{document}

\begin{flushright}
DESY-13-036
\end{flushright}

\vskip 8pt

\begin{center}
{\bf \LARGE {
Quantum Transport and Electroweak Baryogenesis
 }}
\end{center}

\vskip 12pt

\begin{center}
\small
{Thomas Konstandin} \\
--\\
{\em DESY, Notkestr.~85, 22607 Hamburg, Germany } \\
\end{center}

\vskip 20pt

\begin{abstract}
\vskip 3pt
\noindent

We review the mechanism of electroweak baryogenesis. Our focus is on the derivation of quantum transport equations from first principles within the Schwinger--Keldysh formalism. We emphasize the importance of the semiclassical force approach, which provides reliable predictions in most models. In the light of recent electric dipole moment measurements and given the results on new physics searches from collider experiments, the status of electroweak baryogenesis is discussed in a variety of models.

\end{abstract}

\setcounter{tocdepth}{2}

\newpage
\tableofcontents
\newpage

\pagestyle{headings}

\section{Introduction\label{sec:intro}}

The goal of any baryogenesis mechanism is to explain the observed asymmetry between matter and anti-matter
\be
\eta \equiv \frac{n_B - \bar n_B}{n_\gamma} 
\simeq 10^{-10}\, .
\ee
In order to produce such an asymmetry dynamically, several symmetries have to be broken what is summarized by the Sakharov conditions~\cite{Sakharov:1967dj}: Baryon number (B) must not be conserved. Charge conjugation (C) and charge conjugation in combination with parity conjugation (CP) must not be a symmetry. Time reversal must not be symmetry what in the early Universe implies a non-equilibrium state of the plasma. Owing to the Sakharov conditions, baryogenesis is only possible in extensions of the Standard Model (SM). In particular, new sources of CP violation and sizable deviation from thermal equilibrium are essential for a viable baryogenesis mechanism. 

\begin{figure}[t!]
\begin{center}
\includegraphics[width=0.95\textwidth, clip ]{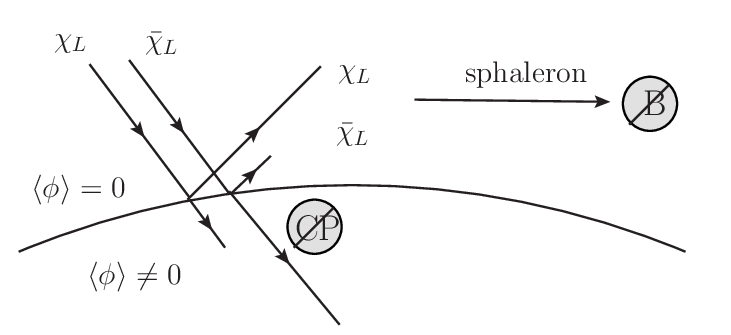}
\end{center}
\caption{
\label{fig:EWBG_sketch}
\small Sketch of the electroweak baryogenesis mechanism: The Higgs bubble walls separate the symmetric from the broken phase. If the reflection of left-handed electroweak particles entails CP violation, the sphaleron process (that only is active in the symmetric phase) generates a net baryon number.  }
\end{figure}

The special appeal of electroweak baryogenesis~\cite{Kuzmin:1985mm} (EWBG) is hereby that only physics of electroweak scales is involved. This makes the scenario in principle testable. The basic picture of electroweak baryogenesis is as follows: At temperatures above the electroweak scale, the electroweak gauge symmetry is unbroken and the Universe is filled with a hot plasma of particles with no net baryon number. The Universe expands and cools and eventually the electroweak gauge symmetry is spontaneously broken via the Higgs mechanism. Electroweak baryogenesis can be realized if this change of phase proceeds by a first-order phase transition. In this case, bubbles nucleate that contain a plasma of broken electroweak symmetry and subsequently expand in the surrounding plasma with unbroken symmetry. Individual particles in the plasma experience the passing bubble interface because of their couplings to the Higgs field. This leads to the reflection of particles and drives the plasma out of equilibrium. Eventually this reflection process entails CP violation and an asymmetry between particles and anti-particles accumulates over time in front of the expanding bubble walls. Since baryon number is conserved up to this point, the opposite CP asymmetry accumulates inside the bubbles of broken plasma. Finally, baryon number is violated due to the sphaleron process that is only active in the unbroken phase. The sphaleron also provides the C violation since it couples only to left-handed particles. This mechanism is most efficient when the particle asymmetries diffuse deep into the unbroken phase where the sphaleron rate is unsuppressed~\cite{Cohen:1994ss}. The mechanism is sketched in Fig.~\ref{fig:EWBG_sketch}.     

Under all models that provide the necessary ingredients for electroweak baryogenesis the minimal supersymmetric standard model (MSSM) has a prominent role. This is mostly due to the fact that the MSSM overcomes (or alleviates) many shortcoming of the SM in some regions of the parameter space: The hierarchy problem of the Standard Model (SM), unification of gauge couplings, the anomaly in the gyromagnetic moment of the muon, viable dark matter candidates and so on. For these reasons, the MSSM is also the most studied framework for electroweak baryogenesis. 

Unfortunately, electroweak baryogenesis is not so easily realized in the MSSM. The main reasons are that the Higgs sector is rather constrained and that CP violation arises in a special form. Even though a strong enough phase transition is possible in a small region of the parameter space (the so-called light stop scenario), the observed baryon asymmetry can only be explained by nearly mass degenerate charginos and/or neutralinos.
Therefore, a reliable analysis of the produced baryon asymmetry has to account for flavor effects as {\em e.g.}~flavor oscillations, resonant enhancements and transport phenomena that are specific to the multi-flavor case. A large part of the literature deals with these complications that are responsible for the large discrepancies in the baryogenesis analysis between different approaches. Recently, the available parameter space for viable MSSM models shrunk significantly with the LHC results, in particular the Higgs searches. All in all, electroweak baryogenesis in the MSSM is technically not ruled out yet, but only possible under rather contrived assumptions and at the cost of additional cancellations and tunings (a more detailed analysis will be given in section \ref{sec:MSSM}). 

The main purpose of the present review is to turn the spotlight on electroweak baryogenesis in models other than the MSSM. The emphasis is hereby on the following aspects:
\begin{itemize}
\item Using the Schwinger-Keldysh formalism~\cite{Schwinger:1960qe,Keldysh:1964ud}, quantum transport equations have been derived in the recent years from first principles in the context of electroweak baryogenesis. Especially, when the CP violation operative in baryogenesis results from the semi-classical force and is not based on flavor mixing, all applied approximations are well justified and allow for robust quantitative predictions. 
\item Recent LHC results marked the discovery of a Higgs-like particle with mass $m_h \simeq 125$ GeV. If this particle is identified with the Higgs particle, this is most relevant for electroweak baryogenesis. The strength of the electroweak phase transition is tightly linked to the Higgs mass. Larger Higgs masses tend to weaken the phase transition and suppress the produced baryon asymmetry. In all models we assume a Higgs mass of above value in this manuscript. 

\item The main motivation for new physics at the electroweak scale (and supersymmetry in particular) comes from the hierarchy problem. The discovery of the Higgs highlights this fact and rules out Higgs-less models as {\em e.g.}~Technicolor. In the last years much progress was made concerning alternative solutions to the hierarchy problem as for example composite Higgs models. These models typically allow for electroweak baryogenesis without much tuning in the Higgs sector.
\end{itemize}
The plan of the review is as follows: In section~\ref{sec:KB} semi-classical transport equations are derived from first principles in the Schwinger-Keldysh formalism. The main result of this section is the Boltzmann equation (\ref{eq:ewbg_one}) that includes a CP-violating semi-classical force at order $\hbar$. Subsequently, in section~\ref{sec:PT} we present how to transit from Boltzmann type transport equations to diffusion equations using the flow ansatz and the complete analysis of the produced baryon asymmetry is illustrated. Finally, in section~\ref{sec:models} the analysis of the baryon asymmetry and its correlation with collider phenomenology is discussed in specific models. The appendix contains the remaining ingredients of the baryogenesis calculation. This includes the characteristics of the phase transition and the sphaleron rate.

\newpage

\section{Quantum kinetic equations\label{sec:KB}}

In this section, we discuss how quantum transport equations can be
derived from first principles in a QFT setting and its application to
electroweak baryogenesis. Main aim of this section is to sketch and
motivate the Schwinger-Keldysh formalism (also known as {\em closed time
path} formalism or {\em in-in} formalism) rather than to discuss it in
complete depth.  The discussion closely follows the derivation in
Ref.~\cite{Kainulainen:2001cn} and the technical review in
refs.~\cite{Prokopec:2003pj, Prokopec:2004ic}. More details can be
found in Refs.~\cite{Calzetta:1986cq, Berges:2004yj, CalzettaHu}; thermal field theory is covered in the books \cite{Kapusta, LeBellac}.

\subsection{The Schwinger-Keldysh formalism}

Starting point of the Schwinger-Keldysh formalism is the observation
that not only scattering amplitudes allow for a representation in
terms of path integrals but also the time evolution of expectation
values of operators~\cite{Schwinger:1960qe, Keldysh:1964ud}. Consider
a quantum mechanical system with coordinate $q$, a basis $n$ and some operator $\hat O$ that at initial time $t_0$ leads to the
matrix elements
\be
\label{eq:mat_ele}
O_{mn}(t_0) = \left< m | \hat O | n \right>. 
\ee
Matrix elements evaluated at later time can be related to
$O_{mn}(t_0)$ via
\be
O_{ab}(t_1) = \sum_{n,m}
\left< a | e^{i \hat H(t_1 - t_0)}|  m \right>
O_{mn}(t_0) 
\left< n| e^{-i \hat H(t_1 - t_0)} | b \right> \, ,
\ee
Hence, unlike scattering amplitudes the time-evolution of a matrix element
involves the evolution of states back and forth in time. 

\begin{figure}[t!]
\begin{center}
\includegraphics[width=0.95\textwidth, clip ]{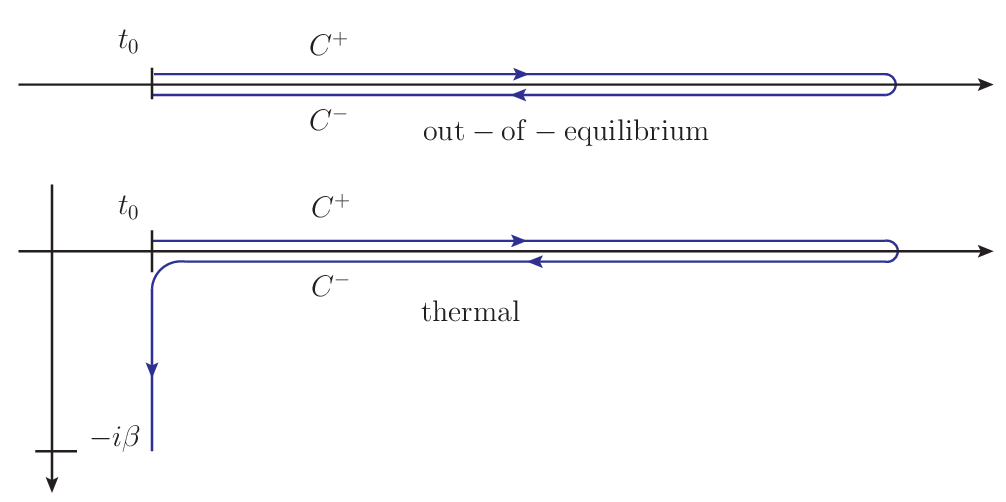}
\end{center}
\caption{
\label{fig:CTP}
\small The closed time path contour for a general out-of-equilibrium system (top) and a system in equilibrium at finite temperature (bottom).  }
\end{figure}

In the path integral formulation, the evolution of the basis states can be expressed as
\be
\left< n | e^{-i \hat H(t_1 - t_0)}|  b \right> 
= \int {\cal D}q \, e^{i \int_{t_0}^{t_1} \, dt \, {\cal L}(q, \dot q)} \, ,
\ee
with the Lagrangian ${\cal L}$ and appropriate boundary conditions.
The time-evolution of an operator can then
be represented as
\be
O_{ab}(t_1) = \int {\cal D}q \,O(t_1)\, e^{i \int_{\cal P} \, dt \, {\cal L}(q, \dot q)} \, ,
\ee
using a closed time path $\cal P$ that goes from $t_0$ to late
times and back, see Fig.~\ref{fig:CTP}. 
It is important to remember
that the two branches of integration are independent such that in the
Hamiltonian picture, operators are path ordered and not time
ordered.   

In QFT the same route can be followed leading to path integrals along the closed time path. As in the quantum mechanics example above the evaluation of operators then leads to path ordered expectation values. This in turn leads to the fact that the Dyson series of
time-dependent perturbation theory does not only involve the
time-ordered Green function but also the anti-time ordered and
unordered ones. This can be expressed efficiently by giving the
two-point functions an additional $2 \times 2$ structure, {\em e.g.}
in the case of a scalar field $\phi$ one defines
\bea
\Delta^{++}(u,v) &\equiv& \Delta^t (u,v)
\equiv -i \left< \Omega | T[ \phi(u) \phi^\dagger(v)] | \Omega \right> \, , \nn \\
\Delta^{+-}(u,v) &\equiv& \Delta^< (u,v)
\equiv -i \left< \Omega |  \phi^\dagger(v)  \phi(u) | \Omega \right> \, , \nn \\
\Delta^{-+}(u,v) &\equiv& \Delta^> (u,v)
\equiv -i \left< \Omega |  \phi(u) \phi^\dagger(v) | \Omega \right> \, , \nn \\
\Delta^{--}(u,v) &\equiv& \Delta^{\bar t} (u,v)
\equiv -i \left< \Omega | \bar T[ \phi(u) \phi^\dagger(v)] | \Omega \right> \, ,
\eea
where $T$ and $\bar T$ denotes time and anti-time ordering,
respectively. Obviously only two of the functions are independent 
and the matrix $\Delta$ in this $\pm$ notation is anti-Hermitian in the sense that 
\be
\Delta^\dagger (u,v) = - \Delta (v,u).
\ee
In many cases it is advantageous to express the two-point functions in
terms of the spectral function ${\cal A} = i (\Delta^> -\Delta^<)/2$
and the symmetric propagator ${\cal F} = (\Delta^> +\Delta^<)/2$. For canonically
normalized fields, the spectral function fulfills the relation 
\be
\label{eq:spec_sum_c}
\left. 2 \partial_{u_0} \, {\cal A}(u,v) \right|_{u_0 \to v_0} = \delta(\vec u- \vec v) \, ,
\ee
that follows immediately from the equal time commutation relations of the field $\phi$.

Ultimately, the matrix elements (\ref{eq:mat_ele}) can be used to
determine the properties of a statistical system using the density
matrix $\hat\rho$
\be
{\rm Tr} \left( \hat\rho \, \hat O \right) = \rho_{mn} O_{nm} \, .
\ee
If the density matrix is known at initial time, operators can be
evaluated at later times using the path integral representation of
$O_{mn}$ as outlined above.  In principle all information about the system can then be inferred from the density matrix.

An alternative way of proceeding is to consider a closed system of
$n$-point functions and to impose the initial conditions on the $n$-point
functions rather than the density matrix. In complete analogy to QFT
calculations, the Schwinger-Dyson equations can be derived from the 2PI 
effective action~\cite{Cornwall:1974vz} in the
non-equilibrium setup. Formally, the equation is the same, namely
\be
\label{eq:SchwingerDyson}
\int d^4w \, ( \square + m^2 + \Pi(u,w) ) \Delta(w,v) = \delta(u - v) \, , 
\ee
where $\Pi$ denotes the self-energy. In a specific model, the
self-energy $\Pi$ can be expressed perturbatively in terms of the
interactions and the two-point functions of the system. This allows to
determine the two-point functions at all times consistently without
resorting to initial conditions in terms of a density 
matrix~\footnote{Strictly speaking, the Schwinger Dyson equation in the 2PI formalism allows only for Gaussian initial conditions. More general initial conditions require the use of the nPI formalism or similar techniques~\cite{Danielewicz:1982kk, Calzetta:1986cq, Berges:2004pu, Borsanyi:2009zza, Garny:2009ni}. In the present context, this problem is not of relevance, since one applies the limit $t_0 \to -\infty$ and hence thermal initial conditions.}.

However, even though these equations are formally the same as the Schwinger-Dyson equations, the two-point functions are understood to have the additional $2\times2$
structure mentioned before. Besides, in many cases statistical systems
are not isotropic or homogeneous such that the two-point functions
$\Delta$ and the self-energy $\Pi$ do not only depend on the relative
coordinate $(u-v)$ but explicitly on both coordinates $u$ and $v$ separately. This feature is
particularly bothersome if the two-point functions are transformed
into Fourier space. Usually Feynman calculus is particularly simple in
Fourier space since the convolutions in coordinate space turn into
conventional products
\be
\int dy \, A(x-y) \, B(y-z) \xrightarrow{F.T.} A(p) \cdot B(p) \, .
\ee
However, if a dependence on the average coordinate remains,
convolutions turn into Moyal star products
\be
\int dy \, A(x,y) \, B(y,z) \xrightarrow{F.T.} A(p,X) \star B(p,X) \, .
\ee
Here $A(p,X)$ denotes the Fourier transform with respect to the relative coordinate $r = (x-y)$ for fixed central coordinate $X = (x+y)/2$
\be
A(p,X) = \int d^4 r \, A(X+r/2, X-r/2) \, e^{i \, r \cdot p} \, ,
\ee
and the Moyal star product is defined using the diamond operator 
\be
\diamond =
\frac12 \left(\overleftarrow{\partial}_p \overrightarrow{\partial}_X -
\overleftarrow{\partial}_X \overrightarrow{\partial}_p\right) \, , 
\ee
by
\be
A(p, X) \star B(p, X) = A(p, X) \,  e^{- i \diamond } \,  B(p, X) \, .
\ee
This representation of two-point functions is called Wigner space and
allows for an interpretation in terms of a semi-classical phase
space. One particularly simple application of this formalism is QFT at finite
temperature what we discuss next.

\subsubsection*{QFT at finite temperature}

The density matrix at finite temperature is given by the Hamiltonian $\hat H$
and the temperature $T=\beta^{-1}$ as
\be
\hat \rho = \exp (- \hat H \beta)\, .
\ee
The partition function of this system
\be
Z = {\rm Tr} \, \hat \rho = \int dq \, \left< q | e^{-\hat H \beta}| q \right> \, ,
\ee
can be represented by extending the closed time path into
the imaginary time direction (see Fig.~\ref{fig:CTP}) and imposing periodic (anti-periodic)
boundary conditions for bosonic (fermionic) fields.
For the two-point functions, the
periodic boundary conditions turn into the Kubo-Martin-Schwinger relation 
\bea
 \left. \Delta^> (u,v) \right|_{u_0 - v_0 = t} &=& 
\left. \Delta^< (u,v) \right|_{u_0 - v_0 = t + i \beta} \nn \\
&&   \xrightarrow{F.T.} \Delta^> (k) = \exp (k_0 \beta)  \Delta^< (k).
\eea
In combination with the spectral sum rule (\ref{eq:spec_sum_c})
\be
\int \frac{dp_0}{2\pi} \, 2 p_0 \, {\cal A}(p) = 1 \, ,
\ee
this yields in equilibrium for a {\em free field}
\bea
{\cal A} (p)
 &=& \pi \delta(p^2 - m^2) \, \sign(p_0) \, ,\nn \\
{\cal F} (p)
 &=& -\pi i \, \delta(p^2 - m^2) \left[ 2 n(|p_0|)+ 1 \right] \, ,
\eea
or equivalently
\bea
\label{eq:Wightman_eq}
\Delta^< &=& - \pi i  \, \delta(p^2 - m^2) \, \sign(p_0) \, n(p_0)  \, , \nn \\
\Delta^> &=& - \pi i  \, \delta(p^2 - m^2) \, \sign(p_0) \, ( n(p_0) + 1 )  \, .
\eea
Here we recover the Bose-Einstein particle distribution function 
\be
n(E) = \frac{1}{\exp(E \beta)-1} \, .
\ee
For particle species that are weakly interacting and close to equilibrium, the spectral function ${\cal A}$ is approximately still given by a $\delta$ function and the corresponding component of the plasma can be described by quasi-particles. The particle distribution function $n(X, p)$ is then encoded in the symmetric propagator ${\cal F}$ or the Wightman functions $\Delta^{<,>}$. 

\subsubsection*{Kadanoff-Baym equations}

It is not surprising that the Wightman functions $\Delta^<$ and
$\Delta^>$ encode the particle densities in the plasma. After all
they represent the particle number operators. This indicates a way to
derive quantum transport equations from first principles: The
Schwinger-Dyson equations (\ref{eq:SchwingerDyson}) in Wigner space (that are also called {\rm Kadanoff-Baym} equations~\cite{KadanoffBaym})
\be
\label{eq:KB_init}
 ( p^2-  m^2 + \Pi(p,X) ) \star \Delta(p,X) = 1 \, , 
\ee
have to be solved with appropriate boundary conditions. In components this equation 
can be brought to the form \cite{Prokopec:2003pj, Prokopec:2004ic}
\be
\label{eq:KB_ini}
( p^2-  m^2 - \Pi^h ) \, \star \, \Delta^{<,>}  
- \Pi^{<,>} \, \star \, \Delta^h = {\rm coll.} \, , 
\ee
where we introduced the collision term
\be
\label{eq:coll_term}
{\rm coll.} = \frac12 \lp \Pi^> \, \star \, \Delta^< - \Pi^< \, \star \, \Delta^>\rp \, ,
\ee
the Hermitian part of the Green function
\be
\Delta^h = \Delta^t - \frac12 (\Delta^< + \Delta^>) \, ,
\ee
and analogous definitions for the self-energy $\Pi$. Once the Wightman functions are known, the particle distribution functions can be read off at late times when the system is again close to equilibrium. According to (\ref{eq:Wightman_eq}) one finds
\bea
n(X^\mu, \vec p) &=& 4 i \, \int_{p_0>0} \frac{dp_0}{2 \pi} \Delta^< \, , \\
1 + \bar n(X^\mu, \vec p) &=& 4 i \, \int_{p_0<0} \frac{dp_0}{2 \pi} \Delta^< \, .
\eea 

Using appropriate boundary conditions, the equations (\ref{eq:KB_ini}) can be readily applied to the problem of electroweak baryogenesis.
Initially the system is close to equilibrium and during baryogenesis driven out of equilibrium. In the case of electroweak baryogenesis this stems from
the bubbles of Higgs vacuum expectation value that give rise to a
space-time dependent mass term $m(X)$. The terms on the left-hand side describe the forces that act on the 
particles and also the diffusion of the particle densities away from the wall.
The term on the right-hand side (that is called {\em collision term}) represents the interactions that drive the system to kinematic and chemical equilibrium. The particle densities of the species under consideration can then be read off from the Wightman functions at late times after the phase transition is completed.

\subsection{Approximation schemes}

In order to make the system of equations (\ref{eq:KB_ini}) more manageable, several
approximations can be applied that we discuss in this
subsection. In the context of electroweak baryogenesis the following
approximations are usually 
employed~\footnote{See Ref.~\cite{Cirigliano:2009yt} for a similar discussion.}:

\begin{itemize}

\item Gradient expansion:

If the background depends only weakly on space and time coordinates, an 
expansion of the Moyal star products in the diamond operator can be performed. Naively, this is a good expansion for electroweak baryogenesis since the background
is only slowly varying in units of the typical momentum scale. To be
specific, in the MSSM the thickness of the Higgs bubble wall is
typically of order $\ell_w \sim 20-30 \, T^{-1}$. At the same time, a typical particle in the plasma has a momentum of order $p \sim T$. Hence, the diamond operator comes with a factor 
$\diamond \sim (\ell_w T)^{-1} \ll 1$.

\item Fluid approximation: 

The plasma is assumed to be close to equilibrium. In particular, it is assumed that two-to-two scatterings (or other interactions that do not change particle numbers) are fast such that the plasma is well described by the local velocity of the different components of the plasma, the local temperatures and the chemical potentials. The particle distribution functions can then be parametrized as
\be
\label{eq:flow_ansatz}
n \simeq \frac{1}{ e^{(u_\mu p^\mu + \mu )\beta} \pm 1 } \, ,
\ee
where $u^\mu$, $\beta$ and $\mu$ are space-time dependent and denote the four-velocity, the inverse temperature and the chemical potential of the components of the plasma. By taking different moments of the transport equations, the equation of motion for these quantities can then be derived (this is exemplified in section \ref{sec:PT}).

\item Weak coupling:

Far away from the source of non-equilibrium the system will reach its chemical equilibrium via interactions that change particle numbers. These interactions are assumed to be slow such that an expansion in the according coupling constants can be performed.

\end{itemize}
In light of these assumptions, one can then simplify the Kadanoff-Baym equations. As a word of caution, notice that the validity of these approximations is not always guaranteed. The prime example are flavor oscillations where the fluid approximation can fail~\cite{Cirigliano:2009yt}. We will comment on this issue in sec.~\ref{sec:sev_flav}.

We will see shortly that the main source that drives the system out-of-equilibrium and induces CP violation arises from a {\rm kinematic} effect that even persist in the limit of vanishing interactions. The deviations from equilibrium are then suppressed by $\epsilon_w \simeq (\ell _w T)^{-1}$ 
while the self-energy is suppressed by coupling constants and loop factors, $\epsilon_{coll} \simeq g^2/4\pi $. 
In particular, the collision term vanishes in equilibrium but also has an explicit factor $\epsilon_{coll}$ from the self-energy. Hence one can neglect the higher gradients in the Moyal star product of the collision term  and write
\be
{\rm coll.} \simeq \Pi^>(p,X) \Delta^< (p,X) - \Pi^<(p,X) \Delta^> (p,X) \, .
\ee
Furthermore, the terms involving the self-energy on the left-hand side of the Kadanoff-Baym equation (\ref{eq:KB_ini}) mostly affect the shape of the spectral function. The term involving $\Pi^h$ renormalizes the mass term while the term involving $\Pi^{<,>}$ leads to a broadening of the spectral function~\cite{Prokopec:2003pj, Prokopec:2004ic}. These terms will also be neglected in the following such that the Kadanoff-Baym equations read
\be
\label{eq:KB_ini2}
( p^2-  m^2 ) \, \star \, \Delta^{<,>}  =  \, \Pi^>(p,X) \Delta^< (p,X) - \Pi^<(p,X) \Delta^> (p,X) \,. 
\ee

\subsection{One bosonic flavor \label{sec:one_bos}}

For a system with only one bosonic degree of freedom, the Wightman functions are
purely imaginary and one can immediately split the Kadanoff-Baym equations into a real 
\be
( p^2-  m^2 )  \cos(\diamond) \Delta^{<,>} (p,X) = 0 \, , 
\ee
and imaginary part
\be
( p^2-  m^2 ) \sin(\diamond) \Delta^{<,>} (p,X) = {\rm coll.} \, . 
\ee
The real part determines the spectral function and is usually called {\em constraint equation}
while the imaginary part describes the variation of the particle distribution functions due to the background 
and is called {\em kinetic equation}. 

To order ${\cal O}(\diamond^2)$ these equations are solved by the ansatz
\be
\Delta^< = 2\pi \delta(p^2 - m^2) \sign(p_0) \,  n(p, X)  \, ,
\ee
where the particle distribution function now fulfills the equation
\be
\label{eq:kin_1f_b}
2\pi \delta(p^2 - m^2) 
\lp 2 p^\mu \partial_\mu n(p, X) + \partial_\mu m^2(X) \partial_{p_\mu} n (p, X)\rp
= {\rm coll.} 
\ee
This equations allow for a simple semi-classical interpretation: Imagine a particle with a space-dependent mass $m^2(z)$ arising from the Higgs bubble and a fixed four-momentum $p^\mu$ in front on the bubble wall. If the particle passes the wall, its mass changes. If the semi-classical particle is on-shell on both sides of the wall, it has to change its four-momentum and the symmetries of the problem dictate that this change arises in $p_z$. This reasoning leads to the relation $p^2_{z,in} + m_{in}^2 = p^2_{z,out} + m_{out}^2$ and the approaching particle perceives the change in mass similar to a potential barrier. In particular, very soft particles cannot fulfill the on-shell condition inside the bubble and are reflected by the bubble wall. If this picture is generalized to a distribution of particles, $n(p,X)$, and a smoothly changing mass profile $m(X)$, this leads to the statement
\be
p^\mu \partial_\mu n(p, X) =  - \partial_\mu m^2(X) \partial_{p_\mu} n (p, X) \, ,
\ee
which is eq.~(\ref{eq:kin_1f_b}) in the absence of interactions. In the language of Boltzmann equations, the change in mass leads to a {\em kinematic effect} that exerts a {\em force} on the particles in the plasma. This effect is purely classical in the sense that it will not be suppressed in the limit $\hbar \to 0$. For electroweak baryogenesis, this effect is interesting since, as we will see, in case of fermions and/or several flavors the kinematic forces can entail CP violation (to first order in $\hbar$). 

Before we do so, let us comment on some additional features of eq.~(\ref{eq:kin_1f_b}) and its solution. First, notice that if the wall is at rest relative to the plasma, the force is absent. In the wall frame the mass depends only on the spatial coordinates, $m(z)$, while in the plasma frame the equilibrium distribution function depends only on the energy, $n(p_0)$. If these two frames coincide the force term $\partial_\mu m^2(X) \partial_{p_\mu} n(p)$ vanishes and the equilibrium solution (with the space-time dependent mass) solves (\ref{eq:kin_1f_b}) everywhere. In terms of particles, soft particles are reflected, $n(\vec p) = n(- \vec p)$, while hard particles replace hard particles on the other side of the bubble wall. During electroweak baryogenesis, deviations from equilibrium are hence additionally suppressed by the wall velocity $v_w$ in case it is substantially smaller than the speed of light. 

Next, notice that the effect persists even in the limit of vanishing interactions. Once the wall is moving, the soft particles are still reflected, $n(\vec p) = n(- \vec p)$, but this is not consistent with the boundary conditions of a plasma moving towards the bubble wall. Also behind the wall the plasma is not in equilibrium. So interactions are essential to establish equilibrium far from the wall but are not so important to generate the out-of-equilibrium situation in the present context. 

Finally, notice that as long as the effect from the wall can be expressed as a force
\be
p^\mu \partial_\mu n(p, X) \, \delta(p^2 -m^2)
= m(X) F_\mu(X) \partial_{p_\mu} n (p, X) \, \delta(p^2 -m^2) \, ,
\ee
the four-current 
\be
J^\mu = \int d^4 p \, p^\mu n(X, p) \, \delta(p^2 -m^2) \, , 
\ee
is conserved
\be
\partial_\mu J^\mu = 0 \, .
\ee
This supports the picture that the effect is kinematic and neither are particles created nor destroyed in the process. Of course, including particle number changing interactions from the collision term modifies this conservation law. On the other hand, energy-momentum 
\be
T^{\mu\nu} = \int d^4 p \, p^\mu p^\nu \,  n(X, p) \, \delta(p^2 -m^2) \, , 
\ee
is not conserved
\be
\partial_\mu T^{\mu\nu} = \int d^4 p \, m \, F^\nu \,  n(X, p) \, \delta(p^2 -m^2) \not= 0 \, ,
\ee
due to the latent heat that is released during the phase transition from the Higgs sector into the plasma. But interactions preserve the (total) energy-momentum tensor 
\be
\int \frac{d^4 p}{(2\pi)^4} \, p^\mu \, {\rm coll.} = 0
\ee
and do not modify this relation (when summed over all species). Ideally, any approximation to the transport equations that is applied subsequently should respect these laws.

\subsection{One fermionic flavor}

In case of a system with one fermionic flavor, the derivation of the Kadanoff-Baym equation parallels the bosonic case. The equation in correspondence to (\ref{eq:KB_ini2}) yields in this case
\bea
\label{eq:KB_ferm}
( \slash \!\!\! p - P_L \, m(X) - P_R \, m^*(X) ) \star S^< (p,X) &=& {\rm coll.} \, ,
\eea
where $S^<$ denotes the fermionic Wightman function. All subleading terms are already neglected and we introduced a complex, space-time dependent mass. Unlike the bosonic case, this equation cannot be simply split into constraint and kinetic equation because the Dirac operator as well as the Green function $S^<$ contain a spinor structure. In the following we assume that the change of the mass is aligned with the momentum of the particle (both in $z$-direction in the following) what makes the problem effectively $1+1$ dimensional. In case these two directions are not aligned, this situation can be achieved by a suitable Lorentz boost~\cite{Kainulainen:2002th}. 

The spinor structure can then be partially decoupled by observing that the Dirac operator commutes with
the following spin operator
\be
\label{eq:spin_op}
S_z = \gamma_0 \gamma_3 \gamma_5 \propto \gamma_1 \gamma_2 \, .
\ee
Using the projectors $P_s = \frac12(\mathbf{1} + s S_z)$, the Dirac operator can be brought to block diagonal form. The block that encodes the vector and axial currents can then be parametrized as
\be
S^< = \sum_{s=\pm} P_s S^<_s \, , \quad
S^<_s = P_s \left[ \gamma_0 g^s_0 + \gamma_3 g^s_3 + g^s_1  + \gamma_5 g^s_2 \right].
\ee
In this notation the $s$-even (odd) parts of $g_0$ encode the vector density (axial $z$-current), $g_3$ encode the vector $z$-current (axial density) and $g_{1/2}$ the scalar/pseudo-scalar ($z$-spin densities).

In the gradient expansion the spinor structure of the Kadanoff-Baym equations can be decoupled~\cite{Kainulainen:2001cn} what leads to the following constraint and kinetic equations for $g_0$
\bea
\label{eq:1fl_con_kin}
\lp  k^2 - |m|^2 - \frac{s}{k_0}|m|^2 \theta^\prime  \rp  g_0^s &=& 0 \, ,  \nn  \\
\lp k_z \partial_z - \frac12 |m^2|^\prime \partial_{k_z} 
- \frac{ s}{2 k_0} (|m|^2 \theta^\prime)^\prime   \partial_{k_z} \rp g_0^s &=& {\rm coll.} \, ,
\eea
where the mass term was parametrized as $m(z) = |m(z)| \exp(\theta(z))$. So the function $g^s_0$ allows again for the ansatz
\be
g_0^s \propto \delta(k_0^2 - \omega_s^2) n_0^s \, , \quad
\omega_s^2 \equiv k_z^2 + |m|^2 + \frac{s}{k_0}|m|^2 \theta^\prime \, ,
\ee
with
\be
\label{eq:ewbg_one}
\lp k_z \partial_z - \frac12 |m^2|^\prime \partial_{k_z} 
- \frac{ s}{2 k_0} (|m|^2 \theta^\prime)^\prime   \partial_{k_z} \rp n_0^s = {\rm coll.} \, .
\ee
The additional CP-violating force in this equation leads to CP-violating deviations from equilibrium in the axial $z$-current. The analogous equation for $g_3$ shows no dependence on the shift in phase $\theta^\prime$. In total, no particles are produced or destroyed. Still, particles with different spins perceive different potential barriers and are reflected differently by the wall. The spin of the particles is hereby conserved while the chirality is not.

If the wall is at rest, $n_0$ does not depend on $k_z$ and away from the wall the particle distribution functions are in their local equilibrium form. The on-shell condition is still different for particles with different spins such that the two-point functions and also the axial current $J_z^5$ depends on the change of phase $\theta^\prime$ in the wall. Since the solution is consistent with the KMS relation, including interactions does not change this picture~\cite{Prokopec:2004ic}. Only if the wall velocity is nonzero the CP violation can diffuse into the symmetric phase and give rise to sizable baryogenesis.

The equation (\ref{eq:ewbg_one}) is the central relation for electroweak baryogenesis with one flavor. The forces on the left-hand side of the equation encode how the plasma is driven out-of-equilibrium and how CP violation manifests itself in the particle densities. The kinetic term in combination with the collision terms dictate how the particle densities diffuse away from the wall. The collision terms also determine how the asymmetries are communicated to the other particle species and finally the weak sphaleron. The complete electroweak baryogenesis calculation in a toy model is sketched in sec.~\ref{sec:PT}.

\subsection{Several flavors \label{sec:sev_flav}}

If several flavors are considered, additional complications arise. The diamond operator comes with a factor $\hbar$ and for one flavor the constraint equation is in leading order algebraic. Besides, the kinetic equation has an overall factor $\hbar$ and is in leading order a classical transport equation. For several flavors, the leading order of the kinetic equation (in the case of bosons) becomes
\be
2 k^\mu \partial_\mu \Delta^<  +  i [m^2 , \Delta^<] - \frac12 
\left\{ m^{2\prime} ,\partial_{k_z} \Delta^<\right\} = {\rm coll.} \, .
\ee
The first two terms of this equation describe flavor oscillations with a frequency $\omega \simeq \Delta m^2/k_z \propto 1/\hbar$, while the third term gives forces similar to what was found in the one flavor case. The Wightman function does encode in the case of several flavors not only semi-classical particle distribution functions but also coherent superpositions of different mass eigenstates. Even though the Wightman function is diagonal in mass eigenbasis far away from the wall, the forces induce off-diagonal entries that participate in the flavor oscillations. This mechanism gives rise to new sources of CP violation. In particular, this effect arises already in leading order in the kinetic equation. In comparison, the semi-classical force found for one flavor contains one more gradient (and hence one more factor $\hbar$). On one hand, this indicates that the flavor mixing effects can be enhanced relative to the semi-classical force. On the other hand, if the oscillation is rather fast, this suppresses the efficient population of any off-diagonal densities. So it is not {\em a priori} clear if the CP violation stemming from mixing or the one from the semi-classical force dominates the produced baryon asymmetry.  

For completeness, we quote the kinetic equation for fermions with several flavors as derived in \cite{Konstandin:2004gy} up to second order in gradients. In this case, it is more appropriate to parametrize  the two-point functions in terms of left-handed and right-handed densities. The equation of the right-handed density reads
\bea
\label{eq:sev_kin}
k_z \partial_z g_R &+& \frac{i}{2} \left[ m^\dagger m, g_R \right]
- \frac14 \left\{ (m^\dagger m)^\prime , \partial_{k_z} g_R \right\} \nn \\
&+& \frac{1}{4k_z} \left( m^{\dagger\prime} m g_R + g_R m^\dagger m^\prime \right) 
- \frac{1}{4 k_z} \left( m^{\dagger\prime} g_L m + m^\dagger g_L m^\prime \right) \nn \\
&-& \frac{i}{16} \left[ (m^\dagger m)^{\prime\prime}, \partial^2_{k_z} g_R \right]
 + \frac{i}{8 k_z} \left[ m^{\dagger\prime} m^\prime, \partial_{k_z} g_R \right] \nn \\
&+& \frac{i}{8} \left( m^{\dagger\prime\prime} m \partial_{k_z} \left(\frac{g_R}{k_z}\right) - 
\partial_{k_z} \left(\frac{g_R}{k_z} \right) m^\dagger m^{\prime\prime}  \right) \nn \\
&-& \frac{i}{8} \left( m^{\dagger\prime\prime} \partial_{k_z} \left(\frac{g_L}{k_z}\right) m  - 
 m^\dagger \partial_{k_z} \left(\frac{g_L}{k_z}\right)  m^{\prime\prime}  \right) \nn \\
&=& {\rm coll.}.
\eea
The corresponding equation for the left-handed density is obtained by the replacements
\be
g_R \leftrightarrow g_L \, \quad m \leftrightarrow m^\dagger \, .
\ee
Notice that this equation does not explicitly depend on the spin quantum number $s$ and we dropped the superscript. The dependence on $s$ appears again when the functions are rewritten in the previous notation via
\be
g^s_L = g^s_0 - s \, g^s_3 \, , \quad g^s_R = g^s_0 + s\, g^s_3 \, , 
\ee
and the lowest order relation $k_z g_3 = k_0 g_0$. Also notice that this kinetic equation does not explicitly depend on the energy $k_0$. Hence, the transport equations for the particle distribution functions can be obtained be integration without knowledge of the spectral function. 

The second term in (\ref{eq:sev_kin}) induces flavor oscillations while the remaining term of first order in gradients are analog to the classical forces in the one flavor case. These terms source the off-diagonal entries (in flavor space) of the Wightman function and contain new sources of CP violation as in the bosonic system with several flavors.
The last two terms reproduce the semi-classical force known from the one flavor case. 

\subsubsection*{Application to the MSSM}

The main application of the equation (\ref{eq:sev_kin}) is chargino (or neutralino) driven electroweak baryogenesis in the MSSM. In this framework, the semi-classical force that drives electroweak baryogenesis in the one flavor scheme is insufficient to account for the observed baryon asymmetry. This is mainly due to a weak phase transition and rather strict constraints from EDM measurements. Hence electroweak baryogenesis in the MSSM has to rely on flavor mixing effects that nominally are suppressed by one less order in the gradient expansion. 

Unfortunately, some of the usual assumptions used in electroweak baryogenesis calculations potentially break down in the case of CP violation stemming from flavor mixing as is discussed in detail in \cite{Cirigliano:2009yt}. The oscillation frequency is of order $\tau_{osc}^{-1} \sim \Delta m^2 / p$ where the $\Delta m^2$ denotes the difference of the mass eigenvalues squared. For soft particles this leads to fast oscillatory behavior and numerically this fast oscillation suppresses the relevance of the off-diagonal entries. On the other hand, flavor  oscillations are important for the new CP-violating terms that arise in the kinetic equations (\ref{eq:sev_kin}) beyond the semi-classical force~\cite{Konstandin:2004gy}. 

If the oscillations are generally assumed to be faster than the background gradients, $\tau_{osc} \ll \ell_w$, the system is in the adiabatic regime~\cite{Cirigliano:2009yt, Cirigliano:2011di}. In the case of the MSSM this seems reasonable since the bubble wall is rather thick, $\ell_w \,T = 10 - 20 $, and the charginos are never mass degenerate in the wall. Hence, the assumption $\tau_{osc} \ll \ell_w$ should be valid for a typical particle in the plasma with $p \sim T$.
In this regime the flow ansatz (\ref{eq:flow_ansatz}) (including a collective oscillation) seems reasonable. Besides, backreactions from the off-diagonal densities on the diagonal ones are small\footnote{Nominally they are second order in gradients and compatible with the semi-classical force terms.} and can be neglected. This is the route followed in~\cite{Konstandin:2005cd}. Unfortunately, the resulting baryon asymmetry is too small to be simultaneously in accord with EDM constraints and the observed baryon asymmetry (a more extensive account of these results is given in sec.~\ref{sec:MSSM}). 

A first study that does not rely on the assumption of fast oscillations was presented in~\cite{Cirigliano:2009yt, Cirigliano:2011di} for a toy model. In this regime, the interplay of off-diagonal and diagonal parts in flavor space is more involved what can lead to a parametric enhancement of CP violation in the diagonal particle densities. In a bosonic toy model, the modes that are most affected by CP violation are the ones where the oscillation frequency is comparable to the background gradients, $\tau_{osc} \sim \ell_w$. As argued before, in the MSSM these particles are rather hard and this leads potentially to a suppression since these hard modes are not very abundant in the plasma. Still, it might turn out that these modes contribute more to the CP-violating particle densities than the bulk of particles in the adiabatic regime. To settle this issue would require an analysis along the lines of~\cite{Cirigliano:2009yt, Cirigliano:2011di} in a fermionic system (namely the chargino sector of the MSSM) which is a daunting task.

\subsection{Other approaches\label{sec:others}}

In this section we briefly discuss to what extent the approach presented in the last section is consistent with other methods found in the literature. In particular we discuss the semi-classical force in the WKB approximation and the mass insertion formalism.

\subsubsection*{Semi-classical force in the WKB approximation}

Historically, the semi-classical force was initially found in the WKB approximation \cite{Joyce:1994fu, Joyce:1994zn, Joyce:1994zt, Joyce:1994kd} and subsequently applied to the MSSM~\cite{Cline:1997vk, Cline:2000kb, Cline:2000nw, Cline:2001rk}. The derivation is a little less clean than the one in the Kadanoff-Baym framework. For example, it relies on the quasi-particle picture what is a stronger requirement than the mere gradient expansion used in the KB approach. 

The derivation goes as follows: Assume again one fermionic particle species with a space-time dependent complex mass term, $m = |m| e^{i \theta}$. The corresponding Lagrangian is 
\be
{\cal L } = \bar \psi \, ( i \slash \!\!\! \partial - P_L \, m - P_R \, m^*) \psi \, .
\ee
Using a local axial transformation, the Lagrangian can be brought to a form where the mass term is real, but an axial gauge field appears
\be
{\cal L } = \bar \psi \, ( i \slash \!\!\! \partial + \gamma_5 \slash \!\!\!\! Z - m) \psi \, ,
\ee
where $Z_\mu = \frac12 \partial_\mu \theta$. Solving the Dirac equation then leads to the dispersion relation of the quasi particles. In the wall frame one finds~\cite{Joyce:1994zn, Joyce:1994zt}
\be
E^2 = p_\bot^2 + 
\lp \sqrt{p_z^2 + m^2} \pm Z_z \rp^2 \, .
\ee
The different signs denote hereby the spin in $z$-direction in the frame with vanishing $p_\bot$ analogue to the construction in the Kadanoff-Baym approach (\ref{eq:spin_op}). The group velocity of the particle is given by 
\be
v_g = \dot z = \frac{\partial E}{\partial p_z} \, ,
\ee
and energy conservation gives the constraint
\be
\dot E = 0 = \dot z \frac{\partial E}{\partial z} 
+ \dot p_z \frac{\partial E}{\partial p_z} \, ,
\ee
and hence $\dot p_z = - \partial_z E$. From these relations the Boltzmann equation can be derived
\be
\frac{dn}{dt}  = \partial_t n + \dot z  \partial_z n + \dot p_z \partial_{p_z} n 
= {\rm coll.} \, .
\ee
Notice that the relation $\dot p_z = - \partial_z E$ ensures that for a static wall the equilibrium particle distribution function (that in this case only depends on energy in the wall frame) is a solution to the Boltzmann equation. 

Let us compare this result with our findings in the Kadanoff-Baym approach.
In the 1+1 dimensional case and for small gradients one finds
\bea
E^2 &=& \lp \sqrt{p_z^2 + m^2} \pm Z_z \rp^2 \nn \\
&\simeq& p_z^2 + m^2 \pm 2 E Z_z 
\eea
Comparing with (\ref{eq:1fl_con_kin}) we see that the force in the WKB approximation is smaller by a factor $m^2/E^2$ what is close to unity for non-relativistic particles. So the result is in rough agreement with the ones later obtained in the Kadanoff-Baym framework. However, the CP-violating term arises through an (axial) gauge transformation what initially lead to some discussion in the literature if this effect is physical. This issue can be resolved by distinguishing between canonical and physical momenta~\cite{Cline:2000nw}.
This careful analysis also recovers the factor $m^2/E^2$ and is then in full agreement with the result from the Kadanoff-Baym framework.

In conclusion, the derivation of the leading order effect in the Kadanoff-Baym framework agrees with the one in the WKB approximation for one fermionic flavor. Nevertheless, the Kadanoff-Baym framework overcame some shortcomings of the semi-classical analysis. First, above ambiguity involving the canonical and physical momenta never arises. Second, the Kadanoff-Baym framework does not assume quasi-particle states from the start. The quasi-particle properties are rather a consequence of the constraint equations to the lowest orders in the gradient expansion. 

\subsubsection*{Mass insertion formalism}

Another approach to CP-violating sources in transport equations is the mass insertion formalism \cite{Huet:1995mm, Huet:1995sh, Riotto:1995hh, Carena:1997gx, Riotto:1997gu, Riotto:1997vy, Riotto:1998zb, Rius:1999zc, Carena:2000id, Carena:2002ss, Lee:2004we, Cirigliano:2006wh}. The formalism has compared the full-fledged Kadanoff-Baym treatment the advantage that it is perturbative what makes even calculations with several flavors straight forward. The main application of this formalism is hence electroweak baryogenesis in the MSSM.

The main idea is to treat the mass term as an interaction and expand the Kadanoff-Baym equations around a plasma with vanishing mass. Formally, the fermionic equivalent of equation (\ref{eq:KB_init}) 
\be
 ( \slash \!\!\! p -  P_L m -P_R m^\dagger - \Sigma(p,X) ) \star \, S(p,X) = 1 \, , 
\ee
is solved perturbatively (neglecting the terms $\Sigma$ arising from 'real' interactions)
\bea
\label{eq:pert_mass}
\slash \!\!\! p \star S^{(1)} = ( P_L m + P_R m^\dagger ) \star S^{(0)} \, ,
\eea
where $S^0$ denotes the equilibrium solution of a massless particle.

On general grounds this formalism gives rise to several objections \cite{Kainulainen:2002sw}:

\begin{itemize}

\item In the case of one flavor, the main effect comes from a shift in the dispersion relation. This effect can only correctly be accounted for if the Kadanoff-Baym equations are solved. In the perturbative picture, the Kadanoff-Baym equations resum an infinite set of diagrams. Even worse, if the operator $\slash \!\!\!p \, \star$ in (\ref{eq:pert_mass}) is inverted one encounters divergences that have to be dealt with. As a simple example, consider the following equation that mimics the constraint equation
\be
(x - a -\Delta a) f(x) = 0\, ,
\ee
with the solution $f(a) \propto \delta(x-a-\Delta a)$. If the equation is expanded in $\Delta a$, one finds $f^{(0)}(a) \propto \delta(x-a)$ and $f^{(1)}(a) = \Delta a \, f^{(0)}/(x-a)$, what is not well defined. The correct behavior can in principle be recovered when one identifies $\delta(x-a)/(x-a) \to - \delta^\prime (x-a)$. However, in the literature on electroweak baryogenesis the problem is usually avoided by introducing a finite width in the spectral function. Potentially, this leads to an overestimation of the effect. Without expanding in $\Delta a$, the result is manifestly finite. 

\item By construction, the resulting Wightman function is local and hence does not contain any transport. To overcome this problem, the resulting deviation from equilibrium is interpreted as a source term and subsequently inserted into a transport equation to make diffusion possible. In the literature different proposals exist how this has to be done, the most plausible being the use of Fick's law \cite{Carena:2002ss}.

\item Flavor oscillations are not correctly reproduced in the studies based on the mass insertion formalism. 

\item Once the source is inserted into the (classical) transport equations, a basis choice has to be made. The observation is that the CP-violating source vanishes in the mass eigenbasis and the interaction eigenbasis is used. On the other hand semi-classical quasi-particles propagate as mass eigenstates what makes this choice questionable. The transport equations obtained in the Kadanoff-Baym framework are in principle  basis independent~\footnote{However, in practice also a basis choice is often made in the Kadanoff-Baym framework when the particle densities are coupled to other species, see e.g.~\cite{Konstandin:2005cd}. So the problem is for the most part only postponed. }.

\end{itemize}

In ref.~\cite{Carena:2002ss} a refined version of the mass insertion formalism was presented. The mass was hereby expanded around a fixed point
\be
m(X) = m(X^0) + (X_\mu - X_\mu^0) \partial_\mu m \, . 
\ee
The derivative term was again treated as an interaction while the mass term was incorporated in the lowest order solution $S^{(0)}$. This overcame some of the problems listed above but also reduced the predicted baryon asymmetry by one order of magnitude.
In this partially resummed form, the main differences between the mass insertion formalism and the Kadanoff-Baym equations seem to be how transport is implemented and the neglect of flavor oscillations. While the Boltzmann type equations arise naturally in the Kadanoff-Baym equations, the mass insertion formalism still requires to use Fick's law or some other classical input to describe transport.

A quantitative comparison between the different approaches in case of the MSSM is given in section \ref{sec:MSSM}.

\newpage

\section{Electroweak baryogenesis: A toy model\label{sec:PT}}

In this section, we connect the analysis of CP-violating particle densities with the explicit calculation of the baryon asymmetry. Namely, we discuss how to transit from Boltzmann to diffusion equations (mostly in the case without flavor mixing). Finally we exemplify the complete calculation in a toy model. Some ingredients, as e.g.~the weak sphaleron rate and the characteristics of the phase transition are covered in the appendices.

\subsection{From Boltzmann to diffusion equations}

To solve the partial differential equations (\ref{eq:ewbg_one}) or (\ref{eq:sev_kin}) is rather demanding without using further approximations. In the following, we discuss only the diffusion equations in models without flavor mixing, where the semi-classical force is the dominant source of CP violation.  

Consider a Boltzmann type equation in the wall frame of the form
\be
\label{eq:boltz}
p_z \partial_z n(\vec p) + m  \, F_z \partial_{p_z} n(\vec p) = {\rm coll.} \, .
\ee
To simplify these partial differential equations further, often the so-called flow ansatz is used. The underlying assumption is that equilibration involves different time scales~\cite{ Berges:2001fi, Berges:2004ce}. When out of equilibrium, the system establishes after a short time kinetic equilibrium by decoherence effects and scattering processes. After this phase, the particle distribution functions of individual species are approximately of the form
\be
\label{eq:flow_ansatz2}
n (\vec p) = \left. \frac{1}{\exp(u_\mu p^\mu + \mu)/T \pm 1} \right|_{p_0=\omega}\, ,
\ee
where $u^\mu$ denotes the plasma four-velocity, $T$ the temperature and $\mu$ the chemical potential. At intermediate times, these quantities are still space-time dependent. Only at later times, the temperature and the four-velocity of the different species equilibrate to each other and the chemical potentials approach an equilibrium consistent with the conserved charges of the system. Similarly, in electroweak baryogenesis the flow ansatz is fulfilled reasonably well everywhere while the correct equilibrium is only attained away from the wall. Furthermore, it is usually also a good assumption in electroweak baryogenesis to use the same temperature for different species. This is owed to the special structure of the CP-violating source\footnote{In the calculation of the wall velocity this would be a poor approximation~\cite{Moore:1995si}.}. 

Before we solve the Boltzmann equations using this ansatz, we discuss in a little bit more detail the connection to the Kadanoff-Baym equations of section \ref{sec:KB}. In the Kadanoff-Baym equations, the distribution functions for anti-particles are given by the negative frequency part using the identification 
\be
\bar n(p^0) = - n(-p^0) \pm 1 \, .  
\ee
Hence anti-particles come in the flow ansatz (\ref{eq:flow_ansatz2}) with the same four-velocity and temperature but opposite chemical potential, as it should be. Of course, in the presence of CP violation small deviations between the chemical potentials and velocities of particles and anti-particles can arise. Another important point is how to connect the CP-violating force to the system of Boltzmann equations. The Boltzmann equations do not contain the full Dirac structure of the Kadanoff-Baym approach but only parametrize the system by four densities of (pseudo-) particles. Typically these are chosen to be left-/right-chiral particles/anti-particles. In contrast in the Kadanoff-Baym approach, spin is a conserved quantum number. In order to translate the semi-classical force (\ref{eq:ewbg_one}) into the chirality basis, the force is transformed into
\be
\label{sec:toy_forces}
m \, F_z \simeq \frac12 |m^2|^\prime \pm \frac{\sign(p_z)}{2 \omega} ( |m^2| \theta^\prime)^\prime \, ,
\ee
where opposite signs apply for left/right-chiral densities and particles/anti-particles respectively.
Strictly speaking this identification is only true for highly-relativistic particles, but we will see below that it reproduces (in leading order in wall velocity) the correct deviation from equilibrium in terms of vector and axial currents.

Using the flow ansatz, different moments of the transport equation (\ref{eq:boltz}) can then be taken in order to reduce the Boltzmann type equation to a diffusion type equation. This leads to the relations
\bea
\label{eq:diff_0}
\lb p_z \rb \, \mu^\prime + \lb p^2_z \rb \, u_z^\prime + \lb m F_z \rb u_z &=& \lb {\rm coll.} \rb \, , \nn \\
\lb p^2_z \rb \, \mu^\prime + \lb p^3_z \rb \, u_z^\prime + \lb p_z \, m F_z \rb u_z &=& \lb {\rm p_z \, coll.} \rb \, , 
\eea
where $u_z$ is in leading order given by the flow of the background (that equals the wall velocity far away from the wall). We used the fact that the flow term and the force fulfill the relations $d\omega/dp_z = p_z/\omega$ and $d\omega/dz = m \, F_z/\omega$ what ensures that the two derivative terms acting on the energy $\omega$ cancel each other~\footnote{In principle there arises an additional term from derivatives acting on the term $\sign{(p_z)}$ but these turn out to be negligible~\cite{Fromme:2006wx}.}. 

The moments are usually defined as
\be
\lb X \rb = \frac{1}{N} \int d^3 p \frac1{\omega} \frac{d n}{d\mu} \, X \, ,
\ee
with a normalization to a fermionic massless degree of freedom in equilibrium
\be
N = \left. \int d^3 p \frac{d n_f}{d\mu} \right|_{m=\mu=u_z=0}.
\ee
In the following we linearize the system in the chemical potentials and the flow velocities. In leading order certain moments are then related by Lorentz boosts, e.g. $\lb p_z \rb \simeq - u_z \, \kappa$ where $\kappa \equiv \lb \omega \rb$ denotes the statistical factor that is 1 (2) for massless fermions (bosons) for a plasma at rest. Furthermore $\lb p^2_z \rb$ is in leading order $1/3$ of the pressure in the plasma and $\lb p^3_z \rb \simeq - 3 u_z \lb p^2_z \omega \rb$.

Next consider the collision terms. The collision integral in the second equation is dominated by elastic scatterings 
\be
\lb p_z \, {\rm coll.} \rb  \simeq - \Gamma^{ela} ( u - \bar u ) \, ,
\ee
(notice that $\Gamma^{ela}$ has dimension three according to this definition). The function $\bar u$ denotes the flow velocity of the background the species mostly scatters with and it is often assumed that this is given by the wall velocity that describes the flow far away from the wall, $\bar u \simeq v_w$. Notice that this approximation is in principle not consistent with the arguments of energy-momentum conservation discussed in sec.~\ref{sec:one_bos}. Still, as long as the background represents a large number of degrees of freedom, this approximation is reasonable.

The collision term in the first equation encodes the particle changing interactions. These are of the form
\be
\lb  {\rm coll.} \rb  \simeq  \Gamma^{inela} \sum_i c_i \mu_i \, ,
\ee
with $c_i$ some integer constants and the subscript $i$ labels the species of the chemical potentials $\mu_i$. One of these interactions constitutes the sphaleron rate that finally biases the baryon number. 
Both sphaleron rates, strong and electroweak, are non-perturbative and cannot be recovered from the collision term as given in (\ref{eq:coll_term}). They have to be added by hand to the network of transport equations.

Finally, consider the forces in the diffusion equation. The CP-conserving force drives the flow of the particles and anti-particles equally away from equilibrium
\be
\lb m \, F_z \rb u_z \simeq \frac12  |m^2|^\prime \lb 1 \rb v_w \equiv S_\mu\, .
\ee
but does not have a large impact on the chemical potentials. The CP-violating force on the other hand contributes mostly to the equation involving the chemical potential
\be
\lb p_z \, m \, F_z \rb u_z \simeq \frac12 (|m^2|\theta^\prime)^\prime \lb |p_z|/\omega \rb v_w \equiv S_u \, .
\ee
In particular, the CP-violating force comes with different signs for the left- and right-chiral fields such that it has only an impact on the axial current as found in the Kadanoff-Baym approach. Besides, it vanishes explicitly for static walls. 

This system of equations can be brought to the form of a diffusion equation by neglecting terms that are second order in the velocities in the second equation. This gives
\be
(u - \bar u) \simeq \frac{1}{\Gamma^{ela}} \lp \lb p_z^2 \rb \mu^\prime + S_u \rp \, .
\ee
Neglecting derivatives acting on the averages and using this in (\ref{eq:diff_0}) yields
\be
D \mu^{\prime\prime} + v_w \kappa \mu^\prime + S_\mu + S_D = \lb {\rm coll.} \rb \, ,
\ee
where we defined the diffusion constant 
\be
\label{eq:diff_consts}
D = \frac{\lb p_z^2 \rb^2 }{\Gamma^{ela}} \, ,
\ee
and the CP-violating source of the form $S_D = S_u^\prime / \Gamma^{ela}$. However, there is no need for these additional approximations and the linear differential equations (\ref{eq:diff_0}) can be easily solved numerically. 

In conclusion, the system of transport equations can after linearization in the velocities and the chemical potentials be brought to the form
\be
\partial_z \Delta J^z_\alpha + \sum_{A,\beta} \Gamma^{inela}_A c^A_\alpha c^A_\beta \mu_\beta = 0 \, ,
\ee
and
\be
\partial_z \Delta T^{zz}_\alpha + \Gamma^{ela}_\alpha (u_\alpha - v_w) = S_\alpha,
\ee
where the indices $\alpha$ and $\beta$ run over all particle species and chiralities and the index $A$ over all interactions. The term $\Delta J^z$ denotes the current of particles minus antiparticles and the expression $\partial_z \Delta J^z$ represents all three terms on the left hand side of the first equation of (\ref{eq:diff_0}). Likewise, the term $\Delta T^{zz}$ denotes the $zz$ component of the energy momentum tensor of the particles minus antiparticles and the expression $\partial_z \Delta T^{zz}_\alpha$ represents the first two terms of the second equation in (\ref{eq:diff_0}). On the other hand, the CP-violating contribution of the force is treated as a source $S_\alpha$. $\Gamma^{ela}_\alpha$ represents elastic scattering rates while $\Gamma^{inela}_A$ stands for the particle number changing interactions that involve the chemical potentials $\mu_\beta$.
The vectors $c^A_\alpha$ represent which particles participate in a specific interaction. 

A conserved current can be represented by a vector $d_\alpha$. In this case all interactions have to preserve the current, $\sum_\alpha c_\alpha d_\alpha = 0$, and the current should be unsourced, $\sum_\alpha d_\alpha S_\alpha= 0$. An example for conserved quantities are electric charge in the broken phase or baryon number if the weak sphaleron process is neglected.

\subsection{A simple diffusion network\label{sec:toy_model}}

In order to determine the final baryon asymmetry, one has to set up a set of transport equations that contains all relevant degrees of freedom. The sphaleron rate will be one of the smallest interaction rates in the this system such that is suffices to neglect backreactions and determine the net baryon number from the left-handed particle density as described in the beginning of appendix~\ref{sec:Gammaws}.

In our toy model the CP-violating source is in the top sector such that we first consider all fast interaction rates involving the tops. These are the Yukawa interactions with the Higgs, the electroweak interactions with the W-bosons and the strong sphaleron rate that involves all quarks. In the broken phase the Higgs vev induces chiral flips between left- and right-handed tops and also Higgs decay into W-bosons.
The relevant particle changing interaction rates are~\cite{Huet:1995sh, Moore:1997im}
\bea
t_L \leftrightarrow t_R + h \, &:& 
\Gamma_y \simeq 4.2 \times 10^{-3} \,  T \, , \nn \\
t_L \leftrightarrow t_R  \, &:& 
\Gamma_m \simeq \frac{m_t^2}{63 T} \, , \nn \\
t_L + b_L + 4 u_L \leftrightarrow t_R + b_R + 4 u_R  \, &:& 
\Gamma_{ss} \simeq 4.9 \times 10^{-4} T \, , \nn \\
h \leftrightarrow 2 W  \, &:& 
\Gamma_h \simeq \frac{m_W^2}{50 T} \, .
\eea
where $u_L$ and $u_R$ collectively denote the left- and right-handed light quarks.

Next, the elastic scattering rates of the Higgs and the top have to specified. These are usually given in terms of the diffusion constants as defined in (\ref{eq:diff_consts}) and calculated in \cite{Joyce:1994zn, Joyce:1994zt, Cline:1994jm, Arnold:2000dr, Arnold:2003zc} and \cite{Cline:2000nw}
\be
D_q \equiv \frac{6}{T} \, , \, 
D_h \equiv \frac{20}{T} \, . 
\ee

The Higgs and W-bosons decay quickly in the broken phase such that neglecting their chemical potential does not have a large impact on the final baryon asymmetry. 
A detailed analysis concerning this point can be found in~\cite{Fromme:2006wx}.
Furthermore, the interactions with the W-bosons are rather fast such that left-handed up and down quarks have similar chemical potentials. The right-handed bottom and the light quarks are only sourced by the strong sphaleron rate and otherwise interact only with very small Yukawa interactions. Hence, the chemical potential of the light right-handed quarks equals the one of the right-handed bottom quark while the light left-handed quarks have the opposite chemical potential. 

Up to this point, the remaining degrees of freedom are the left-handed top and bottom quark with chemical potential $\mu_q$, and the right-handed top and bottom quarks denoted $\mu_t$ and $\mu_b$ respectively. The light right-handed quarks have the same chemical potential as the right-handed bottom quark, $\mu_b$ and the light left-handed quarks the opposite chemical potential. 

Conservation of baryon number then relates these chemical potentials as
\be
(\kappa_t + 1 ) \mu_q + \mu_b + \kappa_t \mu_t = 0 \, .
\ee
The light quarks cancel in this equation since left- and right-handed particles have opposite chemical potentials. We also neglect the bottom masses, $\kappa_b = \kappa_0 =1$. The chemical potential of the right-handed bottom can then be eliminated in the remaining network. For example, the strong sphaleron couples to the combination
\be
\label{eq:ss_combi}
2 \mu_q - \mu_t - 9 \mu_b = ( 9 \kappa_t + 11 ) \mu_q + ( 9 \kappa_t - 1) \mu_t \, .
\ee
In the first line, the term $-9 \mu_b$ represents the nine light quark chiralities including the right-handed bottom. Ultimately, the left-handed baryon chemical potential entering the sphaleron process (see eq.~(\ref{eq:CS_diff}) of appendix \ref{sec:Gammaws}) is given by
\be
\label{eq:ws_combi}
\mu_L = \mu_q - 2 \mu_b = (3 + 2 \kappa_t) \mu_q + 2 \kappa_t \mu_t \, .
\ee
The contribution $-2 \mu_b$ represents the left-handed quarks of the two light families. Notice that if the top is assumed to be light, $\kappa_t=1$, the combination of the chemical potentials that enters the weak sphaleron process (\ref{eq:ss_combi}) is proportional to the combination of chemical potentials that enters the strong sphaleron process (\ref{eq:ws_combi}). Hence, in this limit the final baryon asymmetry is suppressed by the strong sphaleron rate~\cite{Giudice:1993bb}.

We do not quote the full set of equations here. The explicit equations for a network including the Higgs and W-boson fields can {\em e.g.}  be found in refs~\cite{Fromme:2006wx}. The reduced network without Higgs field has been used in ref.~\cite{Bodeker:2004ws} and \cite{Espinosa:2011eu}. A generalization to the two Higgs doublet model is given in ref.~\cite{Fromme:2006cm}. Also the generalization to supersymmetric extensions is extensively discussed in the literature. This includes new damping rates~\cite{Elmfors:1998hh} but also much more complicated diffusion networks. In many cases it is assumed that super-gauge interactions are in equilibrium such that particle species and their superpartners share the same chemical potential. If this assumption is relaxed, the outcome of the diffusion network depends on many more parameters as {\em e.g.} the mass spectrum of all the superpartners. This can lead to very large correction and even to a change in sign in the final baryon asymmetry~\cite{Chung:2008aya, Chung:2009qs, Chung:2009cb}.

In the following, we present some results from \cite{Espinosa:2011eu}. In order to provide the results as model-independent as possible, the source in the top sector has been parametrized via the mass term as
\be
m_t = y_t  \, \phi(z) \, e^{\Theta_t(z)} \, , \nn \\
\ee
using
\bea
\phi(z) &=& \frac{\phi_c}{2} \lp 1 + \tanh(z/\ell_w) \rp \, , \nn \\
\Theta_t(z) &=& \frac{\Delta \Theta_t}{2} \lp 1 + \tanh(z/\ell_w) \rp \, .
\eea
The final asymmetry is then proportional to the change in the top mass phase during the phase transition, $\Delta \Theta_t$. Otherwise, it only depends on the dimensionless quantities $\phi_c/T_c$ and $\ell_w \, T_c$. 

\begin{figure}[t!]
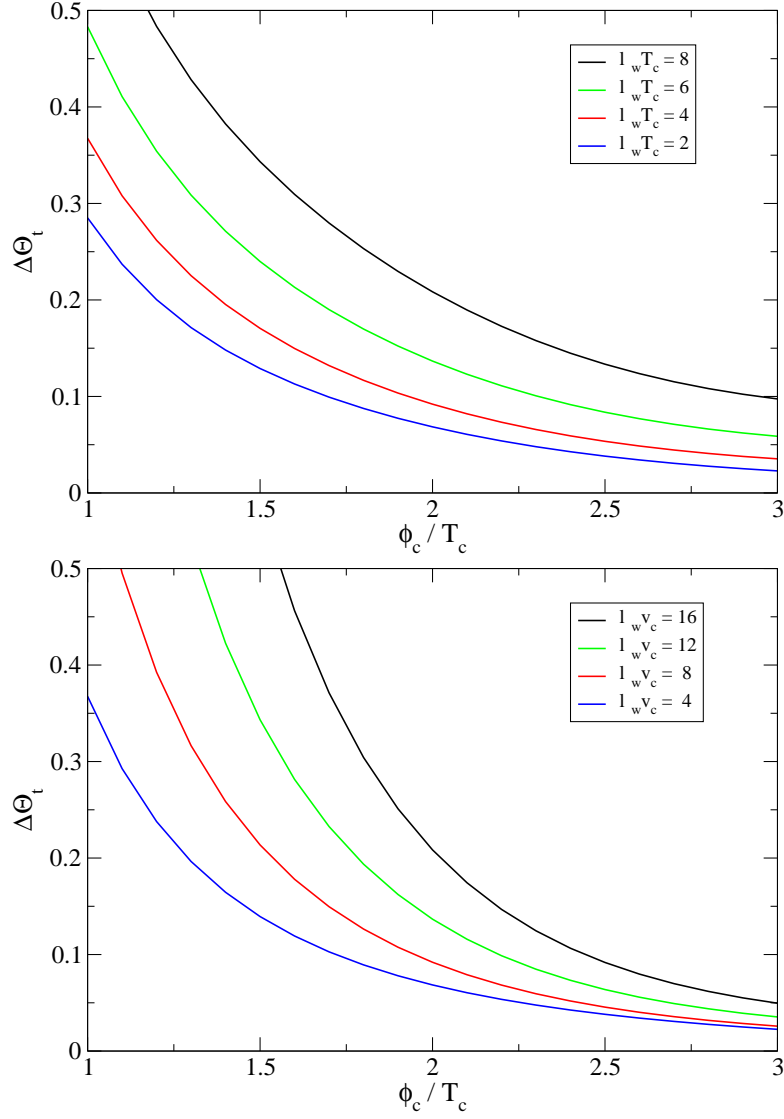

\begin{center}
\includegraphics[width=0.75\textwidth, clip ]{figs/baryo_LT.eps}
\includegraphics[width=0.75\textwidth, clip ]{figs/baryo_Lv.eps}
\end{center}
\caption{
\label{fig:eta_LTLv}
\small The plots show the required change in the top mass phase during the phase transition $\Delta \Theta_t$ in order to reproduce the observed baryon asymmetry. In the upper plot the wall thickness in terms of the temperature is kept constant, while in the bottom plot the wall thickness in terms of the critical vev is kept constant. The plots are adapted from ref.~\cite{Espinosa:2011eu}. }
\end{figure}
Figure \ref{fig:eta_LTLv} shows the required change in the top mass phase during the phase transition $\Delta \Theta_t$ in order to reproduce the observed baryon asymmetry. The baryon asymmetry is very sensitive to the strength of the phase transition, $\phi_c/T_c$. Furthermore, as expected a larger wall thickness reduces the produced asymmetry.
For phase transitions that barely fulfill the baryon washout criterion, $\phi_c \simeq T_c$, a change of phase of order $\Delta \Theta_t \gtrsim 0.3-0.6$ is required for realistic wall thicknesses, $\ell_w \, T_c \simeq 2-8$.

\newpage

\newcommand{\Ztwo}{\mathbf{Z}_2}

\section{Models\label{sec:models}}

The crucial ingredients of electroweak baryogenesis are a strongly first-order phase transition and an appropriate source of CP violation. 

A strong electroweak phase transition is needed for several reasons. First, the nucleated bubbles during the first-order phase transition are the source that drives the plasma locally out of equilibrium and facilitates the establishing of sizable CP-violating currents. Second, the baryon number violating sphaleron processes have to be sufficiently suppressed after the phase transition in order to avoid the washout of the just produced baryon asymmetry. This leads to a constraint on the Higgs vev $\phi_c$ and the phase transition temperature $T_c$ of the form (see sec.~\ref{sec:Gammaws} for a short derivation of this bound)
\be
\label{eq:washout2}
\frac{\phi_c}{T_c} > 1.1 \, .
\ee
In the Standard Model, it is well known that a first-order phase transition is excluded for Higgs masses beyond roughly the W-boson mass \cite{Kajantie:1996mn}. To fulfill the washout criterion (\ref{eq:washout2}), a Higgs mass below $\sim 40$ GeV would even be necessary. This is in contrast to the Higgs mass bound from LEP of $m_H>114$~GeV. Generally, a strong phase transition fulfilling (\ref{eq:washout2}) requires either an extended scalar sector or at least new degrees of freedom that are strongly coupled to the Higgs. 

In electroweak baryogenesis, an appropriate source of CP violation has to be in the form of a complex mass matrix that changes during the phase transition such that a chiral flux is generated close to the bubble wall. This is achieved by coupling the corresponding particles to a vev that constitutes the nucleating bubbles of the phase transition. In many cases this vev arises from the physical Higgs field but more complicated scalar sectors tend to increase the prospects of electroweak baryogenesis. This is because the masses of the Standard Model fermions are proportional to the Higgs vev. Comparison with the sources in (\ref{eq:ewbg_one}) then shows that CP violation is absent. Hence either the masses of the Standard model fermions need to be modified or a new fermionic particle is responsible for the CP-violating flux. In the latter case, the CP-violating flux has to ultimately bias the sphaleron rate. Hence this new degree of freedom is in many models charged under $SU(2)_L$.

At the same time, these new features can leave traces in collider and low energy probes. One major constraint comes hereby from electric dipole moments that constrain new sources of CP violation. Often the induced electric dipole moments arise only at two loop. Still, current bounds on the electron EDM, $d_e< 1.05 \times 10^{-27} \, e \,$cm \cite{Hudson:2011zz}, and neutron EDM, $d_n< 2.9 \times 10^{-26} \, e \,$cm \cite{Harris:1999jx}, heavily constrain realistic models of electroweak baryogenesis. Also the new degrees of freedom responsible for a strong phase transition can have measurable implications. The prime example for this is the Minimal Supersymmetric Standard Model (MSSM) where only light right-handed stops can yield a sufficiently strong phase transition. Such light stops would be copiously produced at the LHC what leads to additional constraints. 

In the following we discuss several models in which electroweak baryogenesis is feasible. We start with relatively simple models with higher dimensional operators and the two Higgs doublet model in which the semi-classical force is operative. Then we discuss the MSSM and its extensions that requires a more sophisticated treatment of CP violation from flavor mixing.

\subsection{The Standard Model with a low cutoff\label{sec:LC1}}

From a bottom-up perspective, the minimal approach to extensions of the Standard Model is to insist on the particle content of the Standard Model and only extend the Lagrangian by higher dimensional operators. Since electroweak baryogenesis requires sizable deviations from the Standard Model at around the weak scale, the suppression of the higher-dimensional operators and the physical cutoff of the theory cannot be much larger in this frame work.
Still, the higher dimensional operators can have an important impact on the phase transition, provide new sources of CP violation and make electroweak baryogenesis a viable option. 

\subsubsection{Phase transition}

The leading operator that modifies the Higgs potential is of the form $(\Phi^\dagger \Phi)^3$, such that the scalar potential of the Higgs vev $\phi$ reads 
\be
\label{eq:LC_Higgspot}
V(\phi) = \mu^2 \phi^2 + \lambda \phi^4 + \frac1{\Lambda^2} \phi^6 \, .
\ee
The new scale $\Lambda$ is the cutoff of the theory where new degrees of freedom become relevant or at least strong coupling phenomena occur. This form of potential can lead to a strong phase transition already in the mean-field approximation where temperature effects only contribute to the quadratic Higgs term, $\Delta V_T \simeq c  \, T^2 \phi^2$. The barrier is then produced by balancing a negative quartic term, $\lambda < 0$, with the positive $\phi^6$ operator~\cite{Grojean:2004xa, Delaunay:2007wb}. The critical temperature is then
\be
T_c^2 = \frac{\Lambda^4 m_H^4 + 2 \Lambda^2 m_H^2 \phi_0^4 - 3 \phi_0^8}{16 \, c \, \Lambda^2 \phi_0^4} \, ,
\ee
where the parameters $\mu$ and $\lambda$ have been expressed in terms of the physical Higgs mass $m_H$ and the observed Higgs vev $\phi_0 \simeq 246$ GeV. The critical vev is given by
\be
\phi_c^2 = \frac32 \phi_0^2 - \frac{m_H^2 \Lambda^2}{2 \phi_0^2} \, .
\ee
There is also an upper limit on $\Lambda$ where the phase transition becomes second order and a lower bound from the fact that the broken phase is the global minimum at $T=0$. As usual an increase in the Higgs mass makes the phase transition weaker. The washout criterion (\ref{eq:washout2}) translates into an upper bound on $\Lambda$. In the full one-loop analysis, the values are~\cite{Delaunay:2007wb}
\bea
\Lambda \lesssim 800 \, {\rm GeV},&& \quad m_H = 125 \, {\rm GeV} \, , \nn \\  
\Lambda \lesssim 900 \, {\rm GeV},&& \quad m_H = 115 \, {\rm GeV} \, .  
\eea
A peculiar feature of the model seems to be that the coefficient of the quartic $\lambda$ is negative. However, a negative quartic can arise quite naturally in effective actions, for example when a heavy scalar is integrated out~\cite{Grojean:2004xa}.

\subsubsection{Electroweak baryogenesis}

Electroweak baryogenesis was considered for this model in~\cite{Bodeker:2004ws}. As an additional efficient source of CP violation served a dimension-six coupling between the Higgs $\Phi$ and the up-quarks
\be
\label{eq:LC_newCP}
{\cal L} \ni \frac{x_{ij}}{\Lambda_{CP}^2} (\Phi^\dagger \Phi) \bar q_i  \, \Phi  \, u_j + h.c. \, ,
\ee
in combination with the usual Yukawa coupling
\be
{\cal L} \ni y_{ij} \bar q_i  \, \Phi  \, u_j + h.c. \, ,
\ee
The resulting fermion masses during the phase transition read
\be
\label{eq:LC_topmass}
m_{ij} = y_{ij} \, \frac{\phi}{\sqrt2} + x_{ij} \frac{\phi^3}{\sqrt8 \Lambda_{CP}^2} \, ,
\ee
what leads to a CP-violating semi-classical force if there are relative complex phases between $y_{ij}$ and $x_{ij}$. The most important effect is in the top sector, since the other quarks are too light to yield a sizable CP-violating flux along the bubble wall. The change of the phase is hence of order 
\be
\label{eq:LC_theta_change}
\Delta \theta \simeq \Im  ( x_t ) \, \frac{\phi^2}{\Lambda_{CP}^2} \, ,
\ee
where $x_t$ denotes the $33$ element of the $x_{ij}$ coupling in the mass eigenbasis of the quarks.

\begin{figure}[t!]
\begin{center}
\includegraphics[angle=-90, width=0.85\textwidth]{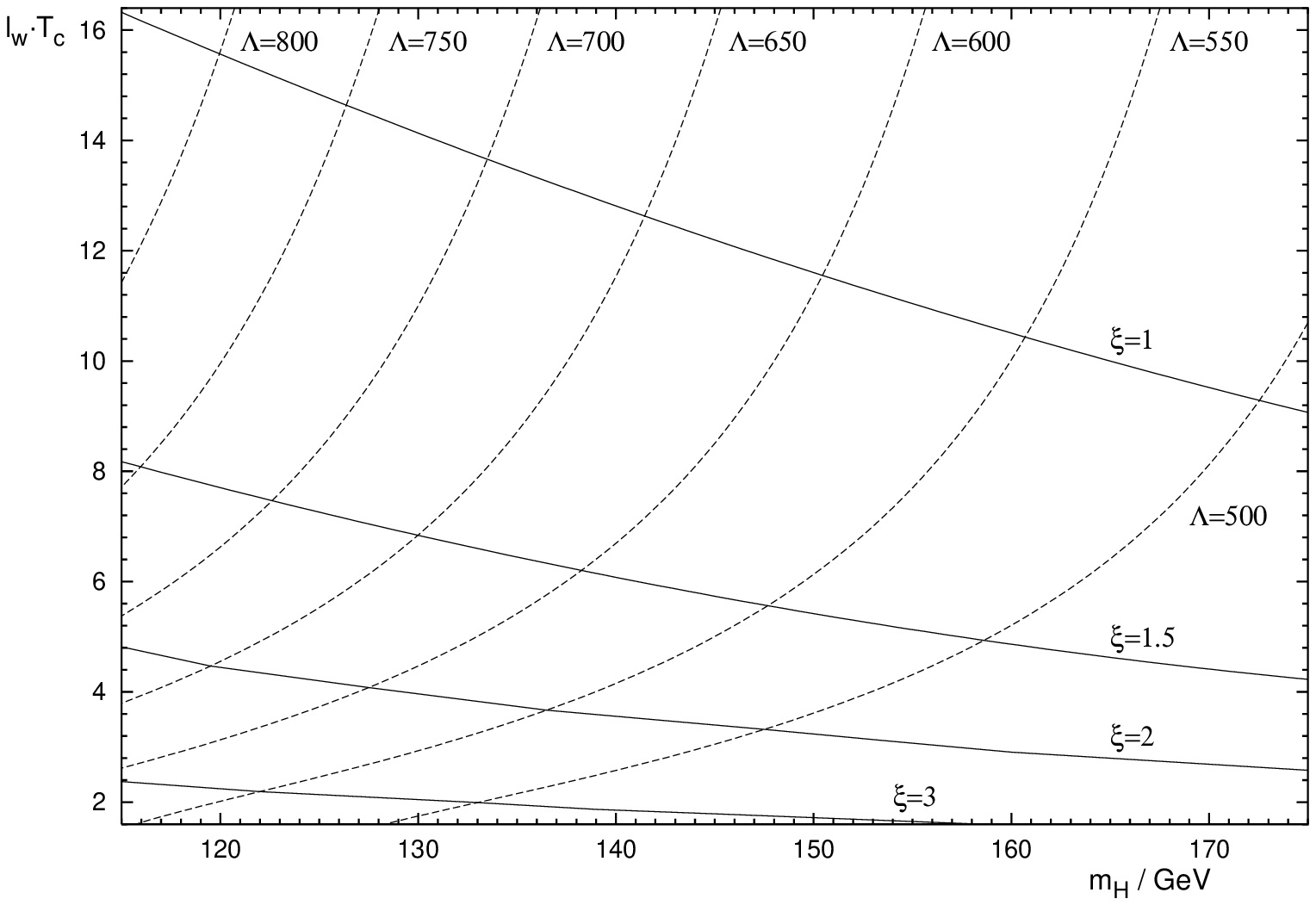}
\end{center}
\caption{
\label{fig:LC_Lw}
\small The wall thickness $\ell_w$ as a function of the Higgs mass. The plot shows also the corresponding values of the scale of new physics $\Lambda$ and the ratio $\xi=\phi_c/T_c$. Plot adapted from \cite{Bodeker:2004ws}.}
\end{figure}
The system of transport equations is the one discussed in section \ref{sec:toy_model}. The only degrees of freedom are the ones from the Standard Model and the dominant source of CP violation is the semi-classical force in the top sector. The only missing ingredient in the present context is the wall thickness. A numerical analysis of all the characteristics of the phase transition (see Fig.~\ref{fig:LC_Lw}) and the analysis of the produced baryon asymmetry is given in ref.~\cite{Bodeker:2004ws}. The final baryon asymmetry is very sensitive to the scale $\Lambda$. The main influence comes hereby from the relation between $\Lambda$ and the critical Higgs vev $\phi_c$. The semi-classical force (\ref{sec:toy_forces}) is proportional to $\phi_c^2$ via the top mass and another factor $\phi_c^2$ stems from the change in the phase (\ref{eq:LC_theta_change}). Besides, the wall thickness $\ell_w$ tends to be smaller for stronger phase transitions and hence lower values of $\Lambda$. For $\Im  ( x_t ) \lesssim 1$, and imposing the observed baryon asymmetry implies the bound $\Lambda_{CP} \simeq \Lambda < 650$ GeV.

\subsubsection{Collider and low energy probes of the model}

Since the model does not contain any new degrees of freedom, no spectacular signatures are expected at colliders. Still, the higher dimensional operators can lead to measurable deviations from the Standard Model. 

In connection to the phase transition, the new operator $\phi^6$ is the essential ingredient. The main collider trace of this new operator is a deviation of the self-couplings of the Higgs in terms of the Higgs mass~\cite{Grojean:2004xa}. The deviations from the Standard model couplings read
\be
\mu = 3 \frac{m_H^2}{\phi_0} + 6 \frac{\phi_0^3}{\Lambda^2} \, , \quad
\eta = 3 \frac{m_H^2}{\phi_0^2} + 36 \frac{\phi_0^2}{\Lambda^2} \, , 
\ee
where $\mu$ ($\eta$) denote the cubic (quartic) self-coupling of the Higgs field. The deviations are pronounced for small Higgs mass, {\em e.g.} $\mu \simeq 2 \mu_{SM}$ for $m_H = 125$ GeV and $\Lambda = 650$ GeV. Still, the discovery of a deviation of this size requires a linear collider~\cite{Grojean:2004xa}. However, in combination with EDM bounds, viable baryogenesis requires an even stronger phase transition what makes even larger deviations in the Higgs sector necessary. This is discussed next.

The new source of CP violation gives potentially much stronger bounds in light of observed limits on flavor changing neutral currents. However, these bounds are more model dependent and in particular hinge on the flavor structure $x_{ij}$ of the new operator (\ref{eq:LC_newCP}). Flavor changing neutral currents potentially arise, because the mass term (\ref{eq:LC_topmass}) is not proportional to the coupling between the Higgs and the fermions
\be
Y_{ij} = y_{ij} \, \frac{1}{\sqrt2} + x_{ij} \frac{3 v^2}{\sqrt8 \Lambda_{CP}^2} \, .
\ee
If the couplings $x_{ij}$ were random numbers of order unity, large deviations in the first two quark families could be observed. For example, the operator (\ref{eq:LC_newCP}) would affect $K - \bar K$ mixing~\cite{Bodeker:2004ws} what implies a bound $\Lambda_{CP} \gtrsim 10^7$ GeV. If on the other hand, $x_{ij}$ has a similar flavor structure as $y_{ij}$, the model is consistent with these constraints as long as $\Lambda_{CP}>500$ GeV. Such a setting is well motivated from the hypothesis of minimal flavor violation \cite{D'Ambrosio:2002ex} and can be achieved in Froggatt-Nielsen type models.

\begin{figure}[t!]
\begin{center}
\includegraphics[width=0.45\textwidth, clip ]{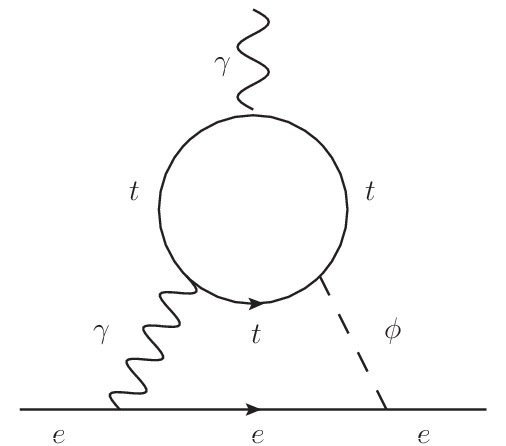}
\end{center}
\caption{
\label{fig:BarrZee}
\small Two loop contribution to the electron EDM of Barr-Zee type.  }
\end{figure}
Constraints from the electric dipole moments are more robust, since they occur even in a one flavor scheme with a relative phase between the couplings $y_t$ and $x_t$. Since this phase is also essential for the production of the baryon asymmetry, this provides a direct link between low energy observables and baryogenesis.
The dominant constraints~\cite{Huber:2006ri} come from the Barr-Zee type contributions to the neutron and electron EDMs (see Fig.~\ref{fig:BarrZee}). For a Higgs mass of $m_H = 125$ GeV this constraint reads $ \Lambda_{CP} \gtrsim \sqrt{\Im (x_t)} \times 750$ GeV and becomes slightly weaker for larger Higgs masses.

\subsubsection{Summary}

Electroweak baryogenesis is a viable possibility in the Standard Model with a low cutoff. The strongest constraints on the model come from the cubic Higgs self-coupling and the upper bounds on the neutron EDM. On general grounds one expects that the new operator in (\ref{eq:LC_Higgspot}) that makes the phase transition strong and the operators that provide the CP violation (\ref{eq:LC_newCP}) are of similar size, $\Lambda \sim \Lambda_{CP}$. This is indeed possible for $\Lambda$ somewhat smaller than $\Lambda_{CP}$. A possible set of parameters is for example
\be
\Lambda \simeq 500 \, {\rm GeV} \, , \quad
\Lambda_{CP} \simeq 1000 \, {\rm GeV} \, , \quad
\Im (x_t) \simeq 1 \, .
\ee
If the Higgs is rather light ($m_H \simeq 125$ GeV), this particular set of parameters will be tested in the near future. The cubic coupling is enhanced by a factor $\sim 3$ compared to the Standard Model what could even be in reach for the high luminosity LHC. Furthermore, the next generation of neutron EDM measurements (assuming an improvement of factor $10$ in sensitivity) can rule out this model of electroweak baryogenesis.

\subsection{Low cutoff: Singlet extension\label{sec:LC2}}

The best motivation for extensions of the Standard Model with a low cutoff comes from composite Higgs models. In composite Higgs models, the light spectrum of the scalar sector depends on the co-set structure of the strongly coupled sector. The degrees of freedom below the scale of strong coupling arise as bound states with pseudo-Goldstone nature. In the minimal model~\cite{Contino:2003ve, Agashe:2004rs, Agashe:2006at} the Higgs is the pseudo-Goldstone boson of the the breaking pattern $SO(5) \to SO(4)$ where $SO(4)$ represents the custodial symmetry of the Higgs sector. In non-minimal models an extended scalar sector appears at low temperatures. In the following we discuss the model with the breaking pattern $SO(6) \to SO(5)$ that serves as a UV completion of the singlet extension of the Standard Model with a low cutoff~\cite{Gripaios:2009pe}.

From a phenomenological point of view, electroweak baryogenesis can be more easily realized in this model than in the Standard Model with a low cutoff. First, the phase transition can be strong already with a renormalizable scalar potential and in the mean-field approximation and does not rely on higher dimensional operators at all. Second, the leading source of CP violation arises already at dimension five. This allows to push the cutoff to a few TeV what is advantageous in view of flavor physics. Last, the dominant contribution to EDM constraints stem from a mixing between the Higgs and the additional singlet degree of freedom. As long as this mixing is small, current constraints from low energy probes are easily fulfilled.

\subsubsection{Phase transition}

As mentioned before, the phase transition can be already strong in mean-field approximation with only renormalizable operators in the scalar potential. Interestingly, this is even true if a $\Ztwo$-symmetry is imposed on the singlet, $s \to -s$. Consider the following potential at the critical temperature:
\be
\left. V \right|_{T=T_c} = \frac{\lambda}{4} \lp \phi^2 + s^2 \phi_c^2/ s_c^2 - \phi_c^2 \rp^2 + \frac\kappa4 \phi^2 s^2 \, .
\ee
The variables $\phi$ and $s$ denote the Higgs and singlet vev and $\phi_c$ and $s_c$ the corresponding values of the vevs in the $SU(2)_L$ and $\Ztwo$-breaking phases at the critical temperature. The first term constitutes a Mexican hat potential with a flat direction that connects the $SU(2)_L$-breaking with the $\Ztwo$-breaking phase. The second term lifts this flat direction and creates a barrier between the two degenerate minima of the potential.

Thermal corrections in the mean-field approximation can be added to this potential via
\be
\Delta V_T = \frac12 (c_\phi \,  \phi^2 + c_s \,  s^2) (T^2 - T_c^2) \, ,
\ee
where the two coefficients $c_\phi$ and $c_s$ read~\cite{Espinosa:2011ax}
\bea
c_\phi &=& \frac1{48} \left[ 9 g^2 + 3 g^{\prime2} + 12 y^2_t + 24 \lambda + 
4 \sqrt{\lambda \lambda_s} + 2 \kappa \right] \, , \nn \\
c_s &=& \frac1{12} \left[ 3\lambda_s + 4 \sqrt{\lambda \lambda_s} + 2 \kappa \right]\, ,
\eea
and we defined $\lambda_s = \lambda \phi_c^4 / s_c^4$. In total the model has four free parameters that can be fixed using the observed Higgs vev $\phi=246$ GeV, the Higgs mass, the singlet mass and the critical temperature. A lower bound on the  singlet mass results from the requirement of a first-order phase transition ($\kappa>0$) while an upper bound on the singlet mass arises from the requirement that the $SU(2)_L$-broken phase is the global minimum at $T=0$. Detailed plots are given in~\cite{Espinosa:2011ax} and also in~\cite{Cline:2012hg}. For fixed Higgs and singlet masses, the critical temperature can always be reduced down to the point where the system becomes very strong, $\phi_c/T_c \sim$ a few. 

In fact, the phase transition proceeds in two stages in this model: At very high temperatures, the singlet vev as well as the Higgs vev vanish and neither the electroweak $SU(2)_L$ nor the $\Ztwo$ symmetry are broken. At lower temperatures the singlet develops a vev that breaks the $\Ztwo$ symmetry. Depending on the parameters, this process can happen at several hundred GeV and is probably rather a cross over than a phase transition. At this stage, domain walls are generated. However, the domain walls are harmless to big bang nucleosynthesis since they disappear in the next stage when the system transits from the $\Ztwo$-breaking phase to the electroweak breaking one.

\subsubsection{Electroweak baryogenesis}

\begin{figure}[t!]
\begin{center}
\includegraphics[width=0.75\textwidth, clip ]{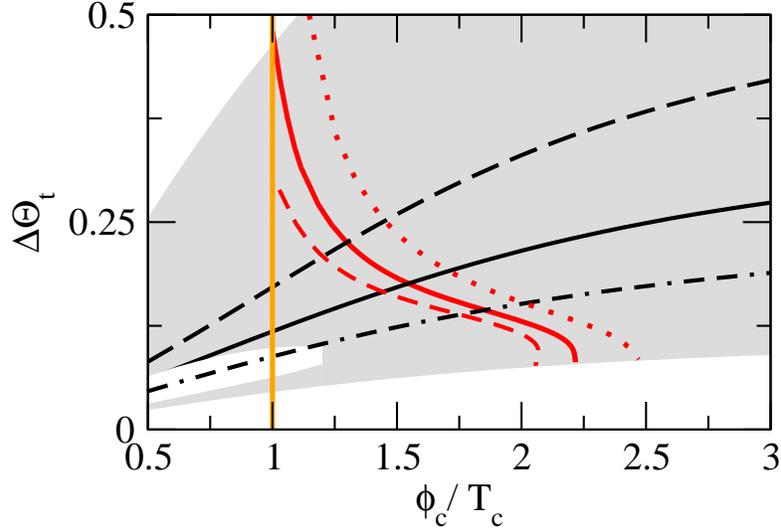}
\end{center}
\caption{
\label{fig:LCS_theta}
\small The shaded region shows possible models in the plane $\xi = \phi_c/T_c$ versus the change in top phase $\Delta \Theta_t$. The Higgs and singlet masses are $m_H = 120$ GeV
and $m_s = $ 130 GeV. The scale of new physics is $\Lambda_{CP} = 500$ GeV. The red lines denote the parameters that reproduce the observed baryon asymmetry. Plot adapted from \cite{Espinosa:2011eu}.  }
\end{figure}

In contrast to the Standard Model with a low cutoff, its singlet extension already has a powerful source of CP violation at dimension five
\be
\label{eq:LCS_newCP}
{\cal L} \ni \frac{x_{ij}}{\Lambda_{CP}} s \bar q_i  \, \Phi  \, u_j + h.c. \, ,
\ee
The resulting fermion masses during the phase transition read
\be
\label{eq:LCS_topmass}
m_{ij} = y_{ij} \, \frac{\phi}{\sqrt2} + x_{ij} \frac{s \, \phi}{\sqrt2 \Lambda_{CP}} \, ,
\ee
what again leads to a CP-violating semi-classical force if there are relative complex phases between $y_{ij}$ and $x_{ij}$. Following the rationale of the Standard Model with low cutoff, we focus on the top sector. The change of the phase of the top mass is of order 
\be
\label{eq:LCS_theta_change}
\Delta \Theta_t \simeq \Im  ( x_t ) \, \frac{s}{\Lambda_{CP}} \, ,
\ee
where $x_t$ denotes again the coupling in the mass eigenbasis of the quarks. Compared to the minimal model with cutoff, the singlet extension has several nice features in view of baryogenesis. First, the phase transition can be rather strong without coming into conflict with a low cutoff. Next, the change of phase (\ref{eq:LCS_theta_change}) is only suppressed by one power of $\Lambda$ what makes baryogenesis in this model easier compatible with a cutoff $\Lambda \sim 2-3$~TeV. With such a high cutoff, it is {\em e.g.}~possible to solve the flavor problem using the 5D GIM mechanism in specific realizations of the composite Higgs mechanism~\cite{Csaki:2008eh}. Furthermore, the singlet vev is in principle expected\footnote{5D realizations of the composite Higgs require a slight tuning to make the electroweak scale and hence the Higgs vev small~\cite{Agashe:2004rs}.} to be larger than the Higgs what further increases the source (\ref{eq:LCS_theta_change}). Some numerical results are shown in Fig.~\ref{fig:LCS_theta}. Electroweak baryogenesis can be viable for $\Delta \Theta_t \gtrsim 1$ what translates into the bound $\Lambda_{CP} <$ a few TeV.

Notice that if the scalar potential is completely $\Ztwo$ symmetric the baryon asymmetry is suppressed. As mentioned above, domain walls are generated at intermediate scales where the singlet vev breaks the $\Ztwo$ spontaneously. At this stage the Universe is divided into regions with positive/negative singlet vev. These regions produce opposite baryon numbers during the electroweak phase transition. In order to avoid this problem, the $\Ztwo$ has to be slightly broken. Already a very small breaking leads to a disappearance of the domain walls and preserves the baryon asymmetry~\cite{Espinosa:2011eu}.

\subsubsection{Collider and low energy probes of the model}

Unlike the Standard Model case, the additional CP-violating operator does not give rise to dangerous flavor observables. First, if the model is approximately $\Ztwo$-symmetric, the operator (\ref{eq:LCS_newCP}) is absent after the electroweak phase transition. Even if the scalar field has a (small) vev after the electroweak phase transition, the Yukawa interactions with the fermions can be diagonalized simultaneously with the fermionic mass terms (\ref{eq:LCS_topmass}) what suppresses flavor changing neutral currents to higher loop order.

\begin{figure}[t!]
\begin{center}
\includegraphics[width=0.55\textwidth, clip ]{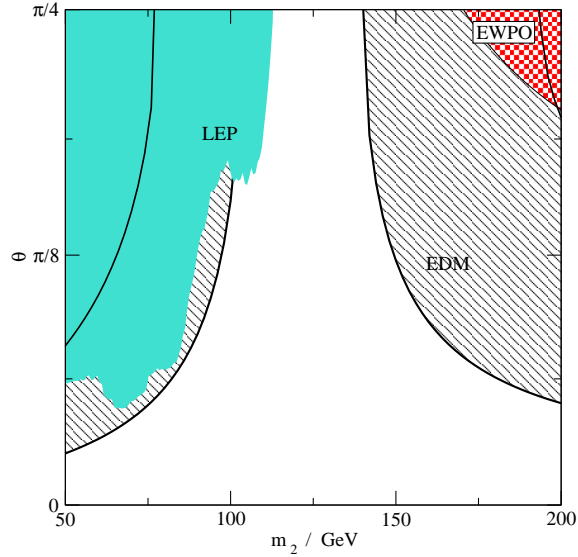}
\end{center}
\caption{
\label{fig:LCS_edm}
\small Bounds on the mass $m_2$ of the mostly-singlet mass
eigenstate, and the mixing angle $\theta$. The mostly-Higgs state has a mass $m_1 = 120$ GeV. The scale of new physics is $\Lambda_{CP} = 500$ GeV. Plot adapted from \cite{Espinosa:2011eu}. }
\end{figure}

In terms of collider traces and electric dipole moments, deviations from the Standard Model arise mostly from a singlet-Higgs mixing. As mentioned above, a very small $\Ztwo$ breaking is required for viable baryogenesis but it is easily compatible with bounds from electroweak precision tests or EDMs as seen in Fig.~\ref{fig:LCS_edm}. 

Another characteristic signal of the model would be a Higgs decay into four fermions via two singlets. If this process can be tested depends however on the coupling of the singlet to the (non-top) fermions and on the mass of the singlet. 

\subsubsection{Summary}

Electroweak baryogenesis can be very easily realized in singlet extensions of the Standard Model with a low cutoff. Since efficient sources of CP violation are present with dimension five operators, the cutoff scale can be slightly larger than in the case of its minimal cousin, $\Lambda<$ a few TeV. Also collider bounds and low energy probes can be easily avoided if the $\Ztwo$ symmetry of the singlet sector is only weakly broken. This makes the model compatible with phenomenology and insensitive to EDM constraints. However, one has to notice that this is somewhat against the philosophy of electroweak baryogenesis that has falsifiability at its core.

\subsection{Two-Higgs-doublet model\label{sec:THD}}

In the two Higgs doublet (THD) model, all necessary ingredients of electroweak baryogenesis are present, even if only renormalizable operators are considered. The most general scalar potential reads
\bea
\label{eq:mo:THD_pot}
V(\Phi_1, \Phi_2) &=& -\mu^2_1 \Phi^\dagger_1 \Phi_1 -\mu^2_2 \Phi^\dagger_2 \Phi_2
- \mu^2_3 \lp e^{i\alpha} \Phi^\dagger_1 \Phi_2 + \, {\rm h.c. } \, \rp \nn \\
&& + \frac12 \lambda_1 (\Phi^\dagger_1 \Phi_1)^2 + \frac12 \lambda_2 (\Phi^\dagger_2 \Phi_2)^2
+ \frac12 \lambda_3 (\Phi^\dagger_2 \Phi_2)(\Phi^\dagger_1 \Phi_1) \nn \\
&& + \lambda_4 | \Phi^\dagger_1 \Phi_2 |^2 
+ \frac12 \lambda_5 \lp ( \Phi^\dagger_1 \Phi_2 )^2 + \, {\rm h.c. } \,  \rp \, . 
\eea
The potential contains two complex (potentially CP-violating) couplings $\mu_3e^{ i \alpha}$ and  $\lambda_5$. Following the conventions of \cite{Fromme:2006cm}, we choose $\lambda_5$ to be real such that $\alpha$ parametrizes CP violation in the scalar sector. As we will see in the next section, the complexity of the scalar potential is also high enough to provide a strong first-order phase transition.

\subsubsection{Phase transition}

In principle there are two regimes in parameter space with a strong first-order phase transition. The first one is similar to the case discussed in the singlet extension of section \ref{sec:LC2}. The phase transition again proceeds in two steps, but unlike in the singlet extension, already this first phase transition breaks the electroweak symmetry in the THD model. This implies that for viable electroweak baryogenesis, this first phase transition has to be strongly first-order, which is not so easily achieved. We hence dismiss this possibility of a two-stage phase transition in the following.

The reason that the phase transition can be much stronger than in the Standard Model is two-fold. The first is that both Higgs doublets acquire a vev after the phase transition and the form of the potential implies that the ratio $\tan \beta$ of these two vevs 
\be
\left< \Phi_1 \right> = ( \, 0, \, h_1 e^{i \Theta_1}\,) \, ,\quad 
\left< \Phi_2 \right> = ( \, 0, \, h_2 e^{i \Theta_2}\,) \, ,\quad 
\tan \beta \equiv h_1/h_2 \, ,
\ee
is not constant during the phase transition. The potential in terms of the vev $\phi^2 = h_1^2 + h_2^2$ is hence not necessarily polynomial and eventually develops a barrier between the two minima at the critical temperature. The second reason is that the scalar potential has enough free parameters to decouple the Higgs mass from the quartic coupling, that in the Standard Model are related by $m^2_h = 2 \lambda \phi_0^2$. It is hence possible to obtain a strong phase transition from the thermal cubic contributions to the effective potential and to satisfy at the same time the LEP bound of $m_h > 114$ GeV.

\begin{figure}[t!]
\begin{center}
\includegraphics[angle=-90, width=0.85\textwidth]{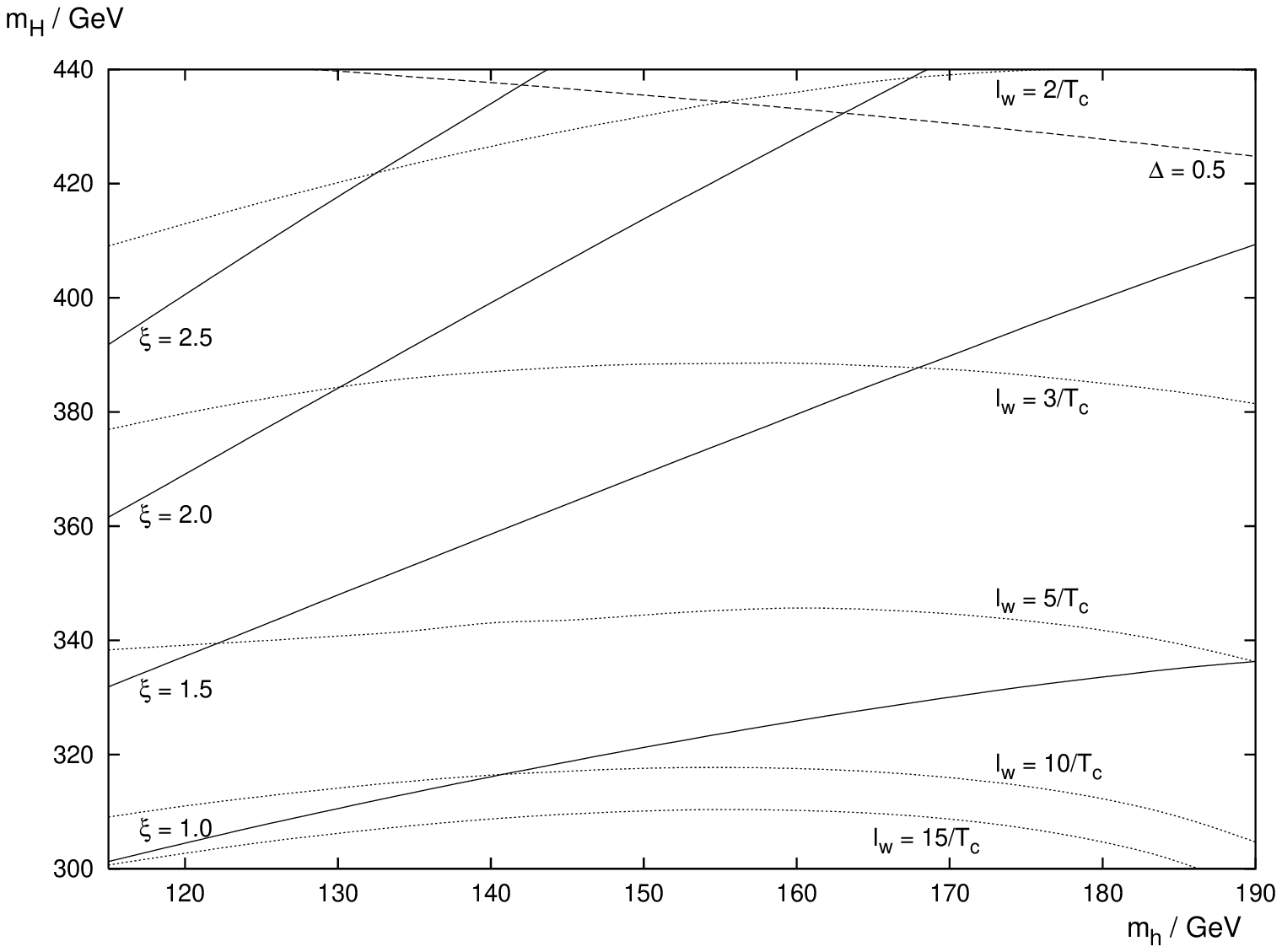}
\end{center}
\caption{
\label{fig:THD_Lw}
\small Lines of constant $\xi=\phi_c/T_c$ and $\ell_w$ as a function of the two scalar masses $m_h$ and $m_H$ and for $\mu_3 = 100$ GeV and $\alpha = 0$. Plot adapted from \cite{Fromme:2006cm}. }
\end{figure}

Overall, relatively strong phase transitions, $\xi = \phi_c/T_c \gtrsim 1.5, \, \ell_w T_c \lesssim 10$, are possible for a Higgs mass above the LEP bound~\cite{Cline:1995dg, Cline:1996mga, Fromme:2006cm, Cline:2013bln}. Some examples are shown in Fig.~\ref{fig:THD_Lw}.

\subsubsection{Electroweak baryogenesis}

The most general THD model with Yukawa couplings of fermions to all two Higgs fields suffers from flavor changing neutral currents already on tree level. To avoid this problem, usually an additional $\Ztwo$ symmetry is invoked that allows to couple the fermions only to one of the two doublets
\be
\Phi_1 \to -\Phi_1 \, , \quad d \to \pm d \, ,
\ee
where depending on the sign in the down sector the THD models type I and type II result, respectively. Notice that the complex phase $\alpha$ in the potential (\ref{eq:mo:THD_pot}) breaks this symmetry explicitly such that electroweak baryogenesis is not possible if this symmetry is also imposed on the scalar sector.

As before, the main source of baryogenesis comes from the top sector and the corresponding Yukawa coupling is of the form
\be
{\cal L} \ni y_t \bar Q_3 \Phi_2 t \, .
\ee
The CP-violating source comes in this model not from the interplay between two operators that both contribute to the top mass, but from the change of the complex phase $\Theta_2$ in the Higgs field that couples to the top
\be
m_t = \frac{y_t}{\sqrt2} h_2 e^{i \Theta_2} \, .
\ee
The change of $\Theta_2$ during the phase transition is hereby induced by the dependence of the scalar potential on the relative phase $\Delta\Theta = (\Theta_1 - \Theta_2)/2$ that arises in the contributions involving $\alpha$.

In \cite{Fromme:2006cm} a part of the parameter space of the THD model is analyzed under the assumption that $\tan \beta$ does not change during the phase transition. However, using this assumption can lead to over-estimating the present CP violation as detailed in \cite{Cline:2011mm}. The reason is the following: The kinetic terms of the Higgs fields coming from the phases yields in the effective action for the vevs the contributions
\bea
S &\ni& \frac12 (\Theta_1^\prime)^2 h_1^2 +  \frac12 (\Theta_2^\prime)^2 h_2^2 \nn \\
&=& \frac12 (\Theta^\prime)^2 (h_1^2 + h_2^2) + \frac12 (\Delta\Theta^\prime)^2 (h_1^2 + h_2^2)
+ \Delta\Theta^\prime \Theta^\prime (h_1^2 - h_2^2) \, ,
\eea
where we defined the average phase $\Theta = (\Theta_1 + \Theta_2)/2$ and the relative phase $\Delta\Theta = (\Theta_1 - \Theta_2)/2$. Since the effective potential does not depend on the average phase $\Theta$, one finds (using the equations of motion)
\be
\Theta^\prime = -\frac{h_1^2 - h_2^2}{h_1^2 + h_2^2} \Delta\Theta^\prime \, .
\ee
Reinserting this into the kinetic term gives
\be
S \ni  (\Delta\Theta^\prime)^2 \frac{h_1^2 h_2^2}{h_1^2 + h_2^2} \, .
\ee
and for the individual phases
\be
\Theta_1^\prime = \frac{2 h_2^2}{h_1^2 + h_2^2} \Delta\Theta^\prime \, , \quad 
\Theta_2^\prime = \frac{-2 h_1^2}{h_1^2 + h_2^2} \Delta\Theta^\prime \, .
\ee
From this it follows that CP violation in the top sector vanishes if one of the vevs vanishes in the trajectory during the phase transition (for $h_2=0$ the top mass vanishes while for $h_1=0$ the phase $\Theta_2$ is constant and no semi-classical force is present). On the other hand, one can always make a basis choice where only one of the Higgs fields has a vev in the broken phase. If a constant $\tan \beta$ was imposed in this basis, CP violation would be absent. So the assumption of constant $\tan \beta$ is not only a basis dependent statement but also of major importance for CP violation. Furthermore, this argument shows that the baryon asymmetry should be suppressed in the limit of very large or very small $\tan \beta$.

Numerically, the study \cite{Fromme:2006cm} found that a baryon asymmetry a few times larger than the observed one is possible in this setup. In contrast, the analysis \cite{Cline:2011mm} additionally implemented (very strict) constraints on $Z \to b \bar b$ and found generically a smaller baryon asymmetry. 

\subsubsection{Collider and low energy probes of the model}

The THD model and its collider phenomenology is widely studied in the literature (for a recent review see \cite{Branco:2011iw}). In the context of electroweak baryogenesis the main signatures are again the electron and neutron EDMs but also the masses of the additional Higgses that have a large impact on the strength of the electroweak phase transition.

The study \cite{Cline:1995dg} found in agreement with \cite{Fromme:2006cm} that for fixed Higgs mass $m_h$ stronger phase transitions can be obtained especially if the additional Higgses are rather heavy. As explained in  \cite{Fromme:2006cm}, this arises from the fact that the larger masses stem from larger quartic couplings and hence corresponds not to a decoupling of the additional Higgses. On the other hand, the quartic couplings are not so essential for collider searches and EDM constraints such that in this limit electroweak baryogenesis is rather unconstrained in the THD model. The limiting factor in this regime is that one wants to preserve perturbativity of the quartic couplings. 

As mentioned before, additional constraints come from $Z \to b \bar b$. The main deviation from the Standard Model stems from the loop contributions of the charged Higgses to this process. In general, this drives the model to larger masses of the charged Higgses and to larger $\tan\beta$. This is problematic for electroweak baryogenesis, since large values of $\tan\beta$ suppress the CP-violating semi-classical force. In \cite{Cline:2011mm} very strict bounds on this process ({\em i.e.} $66\%$ C.L.) have been implemented what has a large impact on the final baryon asymmetry. If this constraint is treated more permissively ({\em e.g.} with $95\%$ C.L.) the corresponding bound is not so relevant and $\tan \beta$ is relatively unconstrained. 

\subsubsection{Summary}

Electroweak baryogenesis is a viable option in the THD model. Without tuning the model allows for a strong first-order phase transition and sufficient CP violation in the scalar sector consistent with EDMs and collider probes. The main disadvantage of the model is that it does not have many benefits beyond electroweak baryogenesis. In particular, the hierarchy problem remains unsolved and flavor issues cannot be solved by a discrete symmetry in the cases where electroweak baryogenesis is  possible.

Over all, an improvement of the measurement of the neutron EDM by a factor around ten can exclude electroweak baryogenesis in the THD model.

\subsection{MSSM\label{sec:MSSM}}

The minimal supersymmetric standard model (MSSM) is one of the most widely studied models today and one of the biggest contenders for the question how the large hierarchy between the electroweak and the Planck scale can be explained. 

The analysis of electroweak baryogenesis is in the MSSM very different compared to other models. First of all, there is no CP violation in the scalar potential and the top sector (beyond the CKM CP violation of the SM), such the CP violation has to arise from a different source than in the cases discussed so far. 
In addition, it is not easy to obtain a strongly first-order phase transition in this setup. In particular, the ratio $\phi_c/T_c$ even in most optimistic scenarios barely fulfills the washout bound (\ref{eq:washout2}) and the wall thickness is rather large, $\ell_w T_c \simeq 20$. This leads to a situation where the semi-classical force falls short to explain the observed baryon asymmetry. Hence electroweak baryogenesis in the MSSM has to be based on a different source of CP violation as {\em e.g.}~the mixing between different charginos (and eventually neutralinos) that can be resonantly enhanced. 

A more extensive recent review of electroweak baryogenesis in the MSSM is given in Ref.~\cite{Morrissey:2012db} and we just present a short overview of the main points here. 

\subsubsection{Phase transition}

The scalar potential in the MSSM is much more constrained than the one of the general THD model. On tree level it reads
\be
V_0 = m_1^2 h_1^2 +  m_2^2 h_2^2 + 2 m_3^2 h_1 h_2 + \frac{g^2 + g^{\prime2}}{8} \lp h_1^2 - h_2^2 \rp^2 \, .
\ee
With this potential the mass of the lightest Higgs bosons is constrained to be below the Z-boson mass. This is not compatible with the bounds from LEP and calls for large one-loop contributions to the Higgs mass 
\be
V_1 = \sum \frac{n_i}{64 \pi^2} m_i^4 \lp \log \frac{m_i^2}{Q^2} - \frac32 \rp \, .
\ee
The dominant contributions to the Higgs mass come hereby from the tops and stops that have Yukawa couplings of order one and the masses, $m_t = y_t h_2$,
\be
{\cal M}_{\tilde t}^2 = \begin{pmatrix}
m_Q^2 + y_t^2 h_2^2 & y_t (A_t h_2 - \mu h_1) \\
y_t (A_t h_2 - \mu h_1) & m_U^2 + y_t^2 h_2^2 \\
\end{pmatrix} \, ,
\ee
where $m_U$, $m_Q$ and $A_t$ are soft supersymmetry breaking terms and $\mu$ stems from a term in the superpotential of form $W \ni \mu \, H_1 \cdot H_2$ . In order to obtain a Higgs mass $m_h \sim 125$ GeV, at least one of the stops has to be rather heavy, $m_{\tilde t_L}> 30$ TeV. This can be achieved by either a large soft mass $m_Q$ or by a large off-diagonal contribution from the $A_t$ term.

\begin{figure}[t!]
\begin{center}
\includegraphics[angle=0, width=0.75\textwidth]{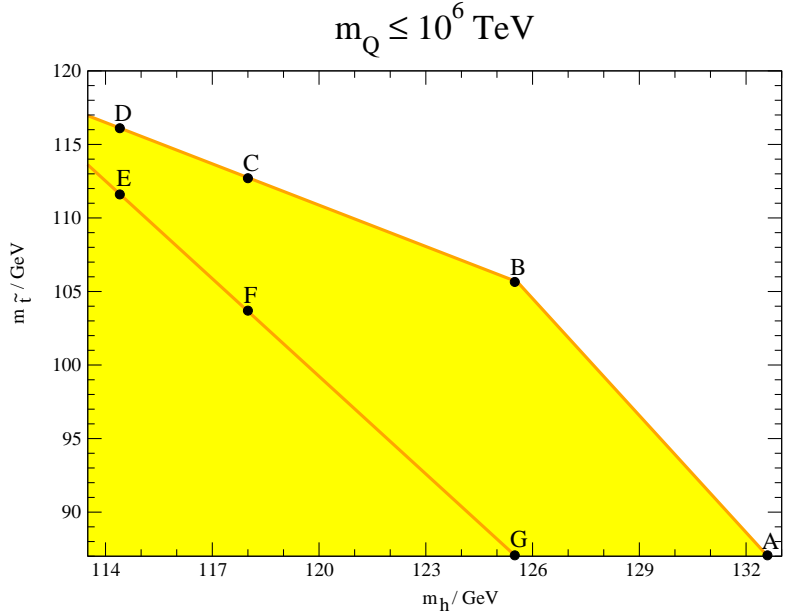}
\end{center}
\caption{
\label{fig:MSSM_PT}
\small The window of strong enough phase transition, $\phi_c / T_c > 1.0$, in the Higgs mass versus light stop mass plane for the MSSM. A strong phase transition and a Higgs mass $m_h \simeq 125$ GeV can only be achieved at the cost of a very heavy left-handed stop, $m_Q \sim 10^6$ TeV. Plot adapted from \cite{Carena:2012np}. }
\end{figure}

The second option is not compatible with a strongly first-order phase transition as we will see in the following. As in the standard model, the potential barrier that is responsible for the first-order phase transition can only arise from thermal cubic terms in the effective potential (see appendix~\ref{sec:PT_in_SM}). Besides the degrees of the freedom of the Standard Model, only the stops can give such a sizable cubic term~\cite{Carena:1996wj, Losada:1998at, Espinosa:1993yi, Carena:2008rt, Carena:2008vj}. This means in turn that the right-handed stop (that is less constrained by electroweak precision tests than its left-handed partner) has to be very light. In particular, a cubic term is only delivered if the mixing between the stops is small and the thermal mass of the right-handed stop is countered by a negative soft mass, {\em i.e.}
\be
m^2_{\tilde t_R}(T) = m_U^2 + y_t^2 h_2^2 + \Pi(T)_{\tilde t_R} \simeq y_t^2 h_2^2 \, .
\ee
Additional constraints arise from the requirement that $\tan \beta$ is not too large and that the stop do not develop a vev at low temperature what would lead to a spontaneous breaking of color. The results of this analysis from \cite{Carena:2012np} is shown in Fig.~\ref{fig:MSSM_PT}. These results also have been qualitatively confirmed in lattice calculations~\cite{Cline:1996cr}.

\subsubsection{Electroweak baryogenesis}

As alluded in section \ref{sec:sev_flav}, the determination of the baryon asymmetry in the MSSM is a controversial topic. One difference to the other models discussed so far is that CP violation does not arise in the top sector. The dominant source of CP violation turns out to be the charginos and neutralinos. For example the chargino mass can be written
\be
{\cal M}_{\chi_\pm} = \begin{pmatrix}
M_2 & g h_2 \\
g h_1 & \mu \\
\end{pmatrix} \, ,
\ee
where $M_2$ and $\mu$ can contain a complex phase.

This mass matrix will lead to a source of the semi-classical force type according to eq.~(\ref{eq:sev_kin}). However, the phase transition in the MSSM is relatively weak~\cite{Moreno:1998bq, Huber:2001xf}, $\phi_c/T_c \simeq 1$, $\ell_w T_c \simeq 20$, such that this source of CP violation is not sufficient to explain the observed baryon asymmetry once EDM constraints are imposed. 

Hence baryogenesis has to be driven by mixing effects in the MSSM. Parametrically, mixing effects are less suppressed because they appear already at first order in gradients as seen in eq.~(\ref{eq:sev_kin}). The determination of the baryon asymmetry based on these mixing effects is to certain extent still an open issue. The mass insertion formalism yields very large baryon asymmetry~\cite{Carena:2000id} but suffers from conceptual problems (see sec.~\ref{sec:others}). Part of these problems can be overcome by resumming Higgs insertions~\cite{Carena:2002ss} but also in this framework some issues concerning finiteness of the results and how transport is established remains. Conceptually the cleanest way to tackle this problem is to use the first principle approach in the Kadanoff-Baym framework. This was done in the analysis \cite{Konstandin:2005cd} that particularly highlighted the importance of flavor oscillations.  But also in this study many simplifying assumptions have been used. Namely, the coherent off-diagonal densities have been assumed to be small. In particular, all contributions that are nominally second order in gradients have been neglected. If these contributions are really small is not so clear since resonant effects can become important when the oscillation length is close to the wall thickness~\cite{Cirigliano:2009yt}. Naively, this resonance condition is for the MSSM charginos only fulfilled for rather hard modes (that are sparse in the plasma) but this does not guarantee that the resonance can give a large enhancement of the baryon asymmetry.

\begin{figure}[t!]
\begin{center}
\includegraphics[angle=0, width=0.48\textwidth]{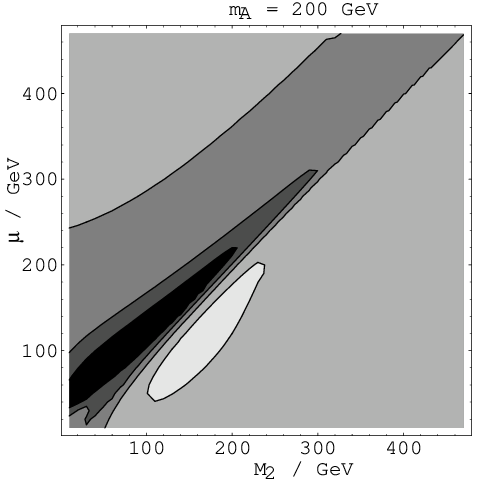}
\includegraphics[angle=0, width=0.48\textwidth]{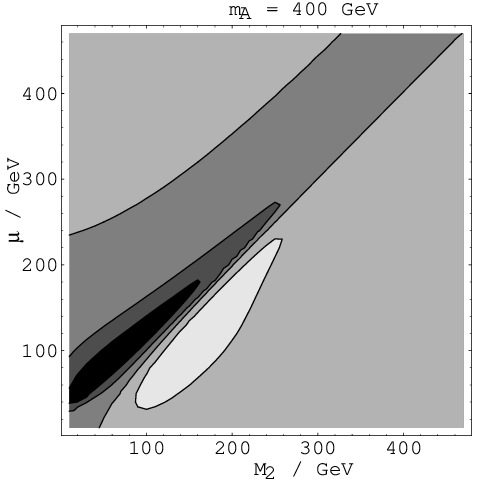}
\end{center}
\caption{
\label{fig:MSSM_eta}
\small Contours of the regions with viable baryogenesis as a function of the two chargino mass parameters $M_2$ and $\mu$. In the black region the baryon asymmetry is larger than observed. Plot adapted from \cite{Konstandin:2005cd}. }
\end{figure}
But there are also some features that are shared by all approaches. For example the baryon asymmetry is suppressed when the charginos are not almost mass degenerate or have a mass much larger than the temperature. This is seen in Fig.~\ref{fig:MSSM_eta} that shows the regions of viable baryogenesis as a function of the two chargino mass parameters. A selection of quantitative results of chargino driven baryogenesis in the MSSM is collected in Table~\ref{tab:MSSM_eta}. 

Beyond theses studies, neutralino \cite{Li:2008ez} or stop driven \cite{Kozaczuk:2012xv} baryogenesis was considered for the MSSM in the literature. Neutralinos have the advantage that they do not suffer from as large EDM constraints as charginos but also are somewhat less efficient in producing the baryon asymmetry~\cite{Li:2008ez}.
\begin{table}[t!]
\begin{center}
\begin{tabular}{|l|p{7 cm}| c |}
\hline
paper & method & $\eta / \eta_{obs}$ \\
\hline \hline
\cite{Carena:2000id} (2000) & 
mass insertion formalism; no Higgs resummation
& $\sim 35 $ \\
\cite{Carena:2002ss} (2002) & 
mass insertion formalism; including Higgs resummation 
& $\sim 10 $ \\
\cite{Lee:2004we} (2004) &
mass insertion formalism; no Higgs resummation; more realistic diffusion network
& $\sim 140$ \\
\cite{Konstandin:2005cd} (2005) &
Kadanoff-Baym formalism; flavor oscillations; assumes the adiabatic regime
& $\sim 3.5$ \\
\hline
\end{tabular}
\end{center}
\caption{
\label{tab:MSSM_eta}
\small The largest possible baryon asymmetry for almost mass degenerate charginos and a maximal CP-violating phase. }
\end{table}

\subsubsection{Collider and low energy probes of the model}

\begin{figure}[t!]
\begin{center}
\includegraphics[angle=0, width=0.48\textwidth]{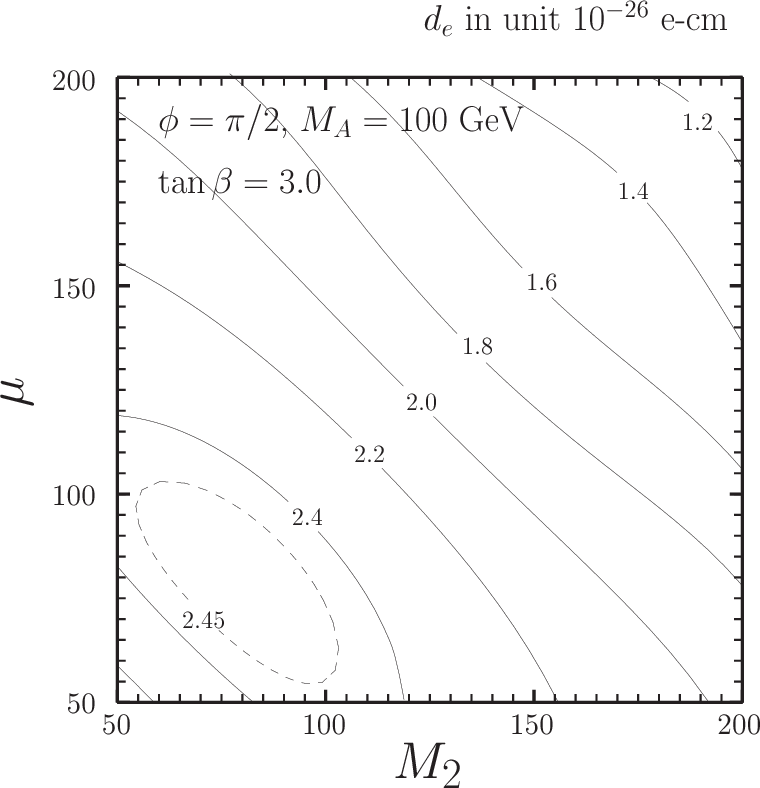}
\includegraphics[angle=0, width=0.48\textwidth]{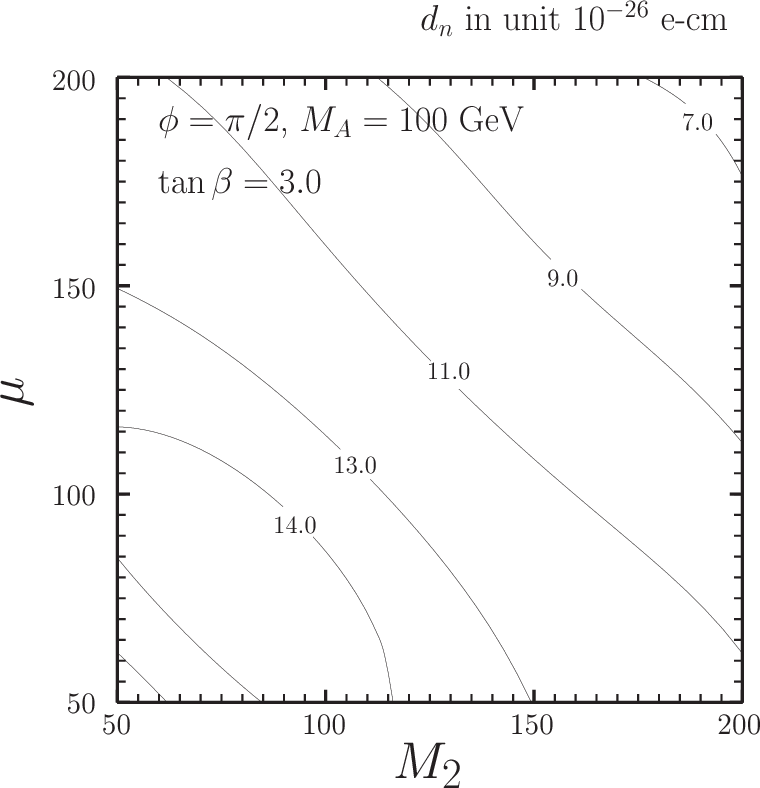}
\end{center}
\caption{
\label{fig:MSSM_EDM}
\small Contours of the electron and neutron EDMs as functions of the two chargino mass parameters and for a maximal CP phase. Plot adapted from \cite{Chang:2002ex}. }
\end{figure}

In the context of electroweak baryogenesis, the MSSM provides some special signatures. The first class of signals comes from the new source of CP violation in the chargino sector. Since the charginos cannot be much heavier than the electroweak scale in electroweak baryogenesis, this leads to Barr-Zee type contributions to the neutron and electron EDMs that are sizable and can be already in conflict with experimental bounds. Furthermore, the dependence of the electron EDM on $\tan \beta$ and the chargino masses is quite similar to the dependence of the baryon asymmetry~\cite{Chang:2002ex, Pilaftsis:2002fe, Cirigliano:2009yd} (see Fig.~\ref{fig:MSSM_EDM}). This implies that the complex phase in the chargino sector cannot be larger than $\arg(\mu^* M_2) \lesssim 0.05$. This excludes chargino driven electroweak baryogenesis in the MSSM in the most conservative approaches (see Table~\ref{tab:MSSM_eta}).

The second class of constraints is connected to the requirement of a strong first-order phase transition. The most severe is hereby the occurrence of stops close to the LEP bound~\cite{Delepine:1996vn, Cline:1998hy, Balazs:2004bu}. More recently, direct searches at LHC are sensitive to light stops such that this is only viable if stop decays are concealed through neutralino states with similar mass or some alternative mechanism~\cite{Carena:2012np}.
Still, the light stops would have a large impact on the Higgs search. In particular, they increase the loop-induced Higgs production rate by gluon fusion by a factor 2 to 3. Besides, light stops lead to a reduced branching ratio for Higgs to di-photons due to a destructive interference with the dominant W-boson loop. Overall, light stops lead to an enhancement of the rate $gg \to H \to VV$ and a slight reduction of the rate $gg \to H \to \gamma\gamma$ compared to the Standard Model. This produces a tension with the current data from Higgs searches~\cite{Menon:2009mz, Cohen:2012zza, Curtin:2012aa} that can be partially relaxed by further assumptions about the particle spectrum~\cite{Carena:2012np}. 

\subsubsection{Summary}

Electroweak baryogenesis in the MSSM is an appealing scenario because the MSSM is the minimal setup that solves the hierarchy problem in a perturbative framework. A Higgs mass of $m_h \sim 125$ GeV produces a tension with minimal supersymmetric models, particularly when a strongly first-order phase transition is demanded. This requires in addition very light stops right above the LEP bound. That these states have been missed at LHC so far is possible but only in case of peculiar masses for the particles that appear in the decay chain of the stops~\cite{Carena:2012np}. Also the EDM constraints are generically in conflict with chargino driven electroweak baryogenesis. So either other sources of CP violation ({\em e.g.}~neutralinos) have to be utilized or the EDMs are small because of a cancellation of different contributions. 

In summary, there remains a region of parameter space in the MSSM in which electroweak baryogenesis is still viable. Still, this possibility appears rather contrived with several requirements arising in different sectors. However, the most constraining requirements can be traced back to the fact that Higgs masses of $m_h \sim 125$ GeV are not easily realized in the MSSM. In extensions of the MSSM where the Higgs mass is achieved more naturally, also the prospects of electroweak baryogenesis are much better. This is explicitly seen in the next model.

\subsection{Next-to-MSSM\label{sec:nMSSM}}

The main aim of singlet extensions of the MSSM is two-fold. First, the $\mu$-problem of the MSSM is solved. This is accomplished by adding a term of form $\lambda S H_1\cdot H_2$ to the superpotential. When the singlet acquires a vev by spontaneous symmetry breaking, this operator produces an effective $\mu$ term. Second, additional contributions to the lightest Higgs mass improve the consistency with current collider constraints. In the following we discuss a variant with only trilinear coupling to the Higgses and a linear term for the singlet in the superpotential as done in \cite{Panagiotakopoulos:2000wp, Menon:2004wv, Huber:2006wf}. More general models can lead to new phenomena as e.g.~transitional CP violation~\cite{Huber:2000mg} or interesting dark matter phenomenology~\cite{Kozaczuk:2013fga}.

\subsubsection{Phase transition}

In this model the scalar potential reads 
\bea
V_0 &=& m_1^2 h_1^2 +  m_2^2 h_2^2 + 2 m_3^2 h_1 h_2 + \frac{g^2 + g^{\prime2}}{8} \lp h_1^2 - h_2^2 \rp^2 \nn \\
&& + \, m_s^2 s^2 + \frac{\lambda}4 h_1^2 h_2^2 + a_\lambda \, s \, h_1 h_2 + t_s \, s \, e^{i \theta_s} + h.c. \, .
\eea
where we defined the vev of the scalar field as $\left< S \right> = s \, e^{i \Theta_s} / \sqrt{2}$. Here, the parameter $\lambda$ results from the term $\lambda S H_1\cdot H_2$ in the superpotential and $t_s$ and $a_\lambda$ are soft SUSY-breaking terms. Of special importance is the contribution $\lambda h_1^2 h_2^2 /4$ which lifts the D-flat direction of the MSSM and can give a sizable contribution to the lightest Higgs mass.

The phase transition can become strong due to the interplay of the singlet and the Higgs vevs and does not rely on thermal loop corrections. Already on tree level the model develops a first-order phase transition when~\cite{Menon:2004wv} 
\be
m_s^2 < \frac{1}{\tilde \lambda} 
\left| \frac{\lambda^2 t_s}{m_s} - \frac{\sin 2\beta}{2} m_s a_\lambda \right| \, , 
\ee
where we defined 
\be
\tilde \lambda^2 \equiv \frac{\lambda^2}{4} \sin^2 (2\beta) + \frac{g^2 + g^{\prime2}}{8} \cos^2 (2\beta) \, .
\ee
For moderate values of $\lambda$, Higgs masses of order $m_h \sim 125$ GeV are possible and consistent with a strong phase transition. However, the parameter $\lambda$ eventually develops a Landau pole at not too high scales what implies the rough bound $\lambda < 0.7$.

\subsubsection{Electroweak baryogenesis}

Electroweak baryogenesis is easier to realize in the nMSSM than in the MSSM for several reasons. First of all, the phase transition can be much stronger. This gives a considerable enhancement in the CP-violating source, that is very sensitive to $\phi_c/T_c$, but also due to a reduced wall thickness. Furthermore, additional complex phases in the parameters $t_s$ and $a_\lambda$ lead to new sources of CP violation. In particular, the phases of the singlet and the Higgs fields changes during the phase transition~\cite{Huber:2006wf, Cheung:2012pg}. The former leads to an additional semi-classical source in the chargino sector via the modified chargino mass matrix
\be
{\cal M}_{\chi_\pm} = \begin{pmatrix}
M_2 & g h_2 e^{i\Theta_2} \\
g h_1 e^{i\Theta_1} & - \lambda  \, s \, e^{i \Theta_s} 
\end{pmatrix} \, ,
\ee
but also to a source in the top sector due to a change in $\Theta_2$ during the phase transition. These contributions arise in the semi-classical force approach and do not rely on mixing. Additional sources by mixing can be as large as in the MSSM but since the semi-classical forces do not require almost mass degenerate charginos these contributions are typically much smaller. This allows for a rather reliable determination of the baryon asymmetry compared to the MSSM.

\subsubsection{Collider and low energy probes of the model}

Compared to the MSSM, collider and EDM constraints are easier to fulfill in the nMSSM. As mentioned before, the lightest Higgs can obtain sizable mass contributions from the coupling to the singlet. However, Higgs masses of $m_h \sim 125$ GeV that rely solely on this coupling lead to a Landau pole in the coupling $\lambda$ below the GUT scale. Hence loop corrections from the stops and tops still have to be sizable and stops heavier than a TeV are required. Notice that light right-handed stops are not essential for a first-order phase transition, such that they can have masses similar to their left-handed counterparts.
 
Constraints from EDM measurements are also easier to avoid than in the MSSM. One reason is that the complex phase in the effective $\mu$ parameter is dynamic. Hence it is possible that the phase is relatively small in the broken phase even though it varied strongly during the phase transition. Also, due to the stronger phase transition, electroweak baryogenesis is more efficient and the observed baryon asymmetry can be reproduced with smaller complex phases in the chargino sector.

\subsubsection{Summary}

In a probabilistic study, the collider and mass constraints provide quite strong bounds on the parameters of the scalar sector. However, once these constraints are passed, a large portion of the remaining parameter space leads to a strong first-order phase transition and viable baryogenesis in the nMSSM~\cite{Huber:2006wf}. In this sense electroweak baryogenesis is a generic feature of the nMSSM.

\subsection{Other models}

For completeness we briefly mentioned in this section other models in which electroweak baryogenesis has been studied. This includes the Beyond MSSM scenario~\cite{Blum:2008ym, Blum:2010by}, the MSSM with an additional $U(1)^\prime$ gauge interaction\cite{Kang:2004pp, Ham:2007wc, Kang:2009rd}, models with R-symmetric supersymmetry~\cite{Kumar:2011np, Fok:2012fb}, the singlet Majoron model~\cite{Cline:2009sn} and left-right symmetric models~\cite{Mohapatra:1992pk}.

\newpage

\section{Conclusions\label{sec:conc}}

The main ingredients of electroweak baryogenesis are a strong first-order phase transition and new sources of CP violation. For this reason, electroweak baryogenesis is ruled out in the SM and heavily constrained in the MSSM. Nevertheless, in models with a more general scalar sector a strong first-order phase transition and electroweak baryogenesis are quite common features. 

From the perspective of electroweak baryogenesis, these models have the added benefit that the determination of the baryon asymmetry is much more robust than in the MSSM. In most of these models, the dominant source of CP violation arises from a semi-classical force that is sensitive to the spin of a single particle species. In contrast, in the MSSM the CP violation operative during the phase transition arises from flavor mixing in the chargino, neutralino or stop sectors. This complicates the analysis through issues that are specific to systems with several flavors as flavor oscillations and resonant enhancements.

Ultimately, whether electroweak baryogenesis is a realistic scenario hinges on the question if and how the hierarchy problem is solved by new physics at the electroweak scale. The LHC discovery of a Higgs-like particle of mass $m=125$ GeV indicates that the MSSM can only solve the hierarchy problem at the cost of introducing a little hierarchy problem. This makes models with extended scalar sectors very attractive and in turn electroweak baryogenesis a promising mechanism for explaining the observed baryon asymmetry of the Universe.

\section*{Acknowledgments}

It is a pleasure to thank my collaborators that worked with me on electroweak baryogenesis and related topics, in particular Stephan Huber, Tomislav Prokopec and Michael~G.~Schmidt. I also would like to thank Valery Rubakov for motivating me to write this review and Mathias Garny for carefully reading the manuscript and his helpful suggestions. Finally, I would like to thank the corresponding authors for their permission to reproduce figures.

\newpage

\appendix

\section{The weak sphaleron rate
\label{sec:Gammaws}}

One essential ingredient of electroweak baryogenesis is the weak sphaleron rate~\footnote{An early review on the sphaleron rate in the context of electroweak baryogenesis is given in~\cite{Rubakov:1996vz}.}.
The sphaleron is a static solution to the field equations of the electroweak sector of the Standard Model. This configuration is a saddle point of the electroweak potential energy and quantifies the dominant baryon and lepton number violating processes at finite temperature in the early universe. 
It couples to the left-handed fermions and anti-fermions of the Standard Model and equally violates lepton and baryon number.  
In the presence of a (eventually local) CP-asymmetry in the left-handed particle densities, the sphaleron is biased towards a net baryon number. At the same time any pre-existing baryon number diffuses as long as baryon minus lepton number is conserved, $B=L$. The baryon asymmetry obeys the equation~\cite{Cline:1997vk} 
\be
\label{eq:CS_diff}
v_w \frac{dn_B}{dz} = \frac32 \Gamma_{ws} 
\left( \frac{\mu_L}{T} - \frac{15}{2} \frac{n_B}{T^3} \right) \, ,
\ee
where $\Gamma_{ws}$ is the weak sphaleron diffusion rate and $\mu_L$ denotes the chemical potential of the left-handed fermions. The final baryon asymmetry is then given by integration
\be
\eta = \frac{n_B}{s} = \frac{405 \Gamma_{ws}}{4 \pi^2 v_w g_* T^4}
\int_0^\infty dz \, \mu_L \, e^{-\nu z} \, ,
\ee
with $g_* \simeq 106.75$ the effective number of degrees of freedom at electroweak temperatures and we defined $\nu \equiv 45 \Gamma_{ws}/ 4v_w T^3$. The chemical potential $\mu_L$ falls off at least as $e^{- D_q z}$ in the symmetric phase where $D_q$ is the quark diffusion constant. Thus for large wall velocities $v_w$, the exponent $-\nu z$ is irrelevant and the dependence on the wall velocity is inherited from the chemical potential $\mu_L$ that is in leading order linear in $v_w$. Hence for $\nu \ll D_q$ and $v_w \ll 1$ the final baryon asymmetry depends only weakly on the wall velocity. If $v_w$ approaches the speed of sound, $c_s = 1/\sqrt3$, diffusion should become inefficient (which however is not correctly reproduced in the analysis of sec.~\ref{sec:PT} that assumes small wall velocities). In the limit of very small wall velocities the exponent becomes important and leads to further suppression. This indicates that the wall is so slow that the sphaleron is saturated. In this regime backreactions on the left-handed chemical potential $\mu_L$ should not be neglected.

On one hand, the sphaleron rate has to be large during the phase transition in the symmetric phase in front of the wall. The CP violation in the reflection of particles leads to a net CP-violating particle density in front of the wall. If this particle density carries (positive) lepton or (negative) baryon number, the sphaleron process is biased towards a positive net baryon number. In order to produce a baryon asymmetry of the observed magnitude $\eta \simeq 10^{-10}$ the sphaleron process should be considerably larger than $\eta$ in electroweak units. 

This sphaleron rate in the symmetric phase was controversially discussed for some time in the literature~\footnote{A nice summary of the {\em status quo} can be found in the talk~\cite{Moore:2000ara}.}. The main problem is that the sphaleron rate is non-perturbative due to the large occupation number of soft modes but also sensitive to the dynamics of the hard modes in the plasma~\cite{Arnold:1996dy}. The system is successfully described by B\"odekers effective theory~\cite{Bodeker:1998hm, Bodeker:1999gx, Bodeker:1999ey} that can be easily simulated on a lattice. In conclusion, the weak sphaleron rate in the symmetric phase reads
\be
\Gamma_{ws} = \kappa \, \lp \frac{g_w^2 T^2}{m_D^2} \rp
\alpha_w^5 T^4 \, ,
\ee
where $m^2_D = \frac{11}{6} g_w^2 T^2$ is the Debye mass of the weak gauge fields and $g_w$ is the gauge coupling of the weak interactions. Numerically the coefficient $\kappa$ is given by $\kappa \simeq 40$. Including the dynamics of the Higgs field slightly reduces this number and one finds~\cite{Moore:2000mx}
\be
\Gamma_{ws} \simeq 1.0 \times 10^{-6} \, T^4 \, .
\ee
This is in principle sufficiently fast for electroweak baryogenesis.

On the other hand, the sphaleron rate in the broken phase should be smaller than in the symmetric one. For equal sphaleron rate no net baryon number would be generated during the phase transition, since the plasma in the bubble carries the opposite lepton and baryon number densities compared to the plasma in front of the wall. In fact, the sphaleron rate in the broken phase must be many orders of magnitude smaller than the rate in the symmetric phase. After the phase transition, the plasma components inside and outside the Higgs bubbles mingle again. Even though a net baryon (and equal lepton) number was generated during the phase transition, the real equilibrium state of the system is still $B=L=0$. If the sphaleron process is still active after the phase transition, the system has a time of order of the Hubble scale to attain this equilibrium. Hence, in order for electroweak baryogenesis to work, the sphaleron rate must be slow compared to the Hubble expansion.

The sphaleron rate in the broken phase is accessible to semi-classical analysis~\cite{Klinkhamer:1984di, Arnold:1987zg} and is exponentially suppressed by the sphaleron energy
\be
\Gamma_{ws} \simeq T^4 \, e^{- E_{sp}/T} \, .
\ee
The sphaleron energy is proportion to~\cite{Klinkhamer:1984di} 
\be
E_{sp} \simeq \frac{4 \pi \phi_c}{g_w} \, \Xi \, ,
\ee
and numerically one finds $\Xi \simeq 2.8$. If one requires that the sphaleron rate is slow compared to Hubble expansion, $\Gamma_{ws} \ll H T^3$, this leads to~\cite{Arnold:1987zg, Shaposhnikov:1987tw, Shaposhnikov:1987pf, Arnold:1987mh}
\be
\label{eq:washout}
\phi_c \gtrsim 1.1 \, T_c \, ,
\ee
This is the so-called sphaleron washout criterion\footnote{For a more detailed discussion of this argument see also~\cite{Patel:2011th, Funakubo:2009eg}.}.

Also the sphaleron rate in the broken phase has been confirmed non-perturbatively on the lattice~\cite{Moore:1998swa}. Recently, the first lattice calculations connecting the symmetric phase with the broken phase have been presented~\cite{D'Onofrio:2012jk}, confirming the picture developed in the two different phases in a unifying framework.

\section{Semi-classical approach to phase transitions
\label{sec:app_tunnel}}

The formalism to describe semi-classical tunneling was pioneered in condensed matter systems by Langer \cite{Langer:1969bc}, in quantum field theory by Coleman \cite{Coleman:1977py, Callan:1977pt} and at finite temperature by Linde \cite{Linde:1980tt}. A review of the topic can be found in~\cite{Quiros:1994dr}.

\begin{figure}[t!]
\begin{center}
\includegraphics[width=0.65\textwidth, clip ]{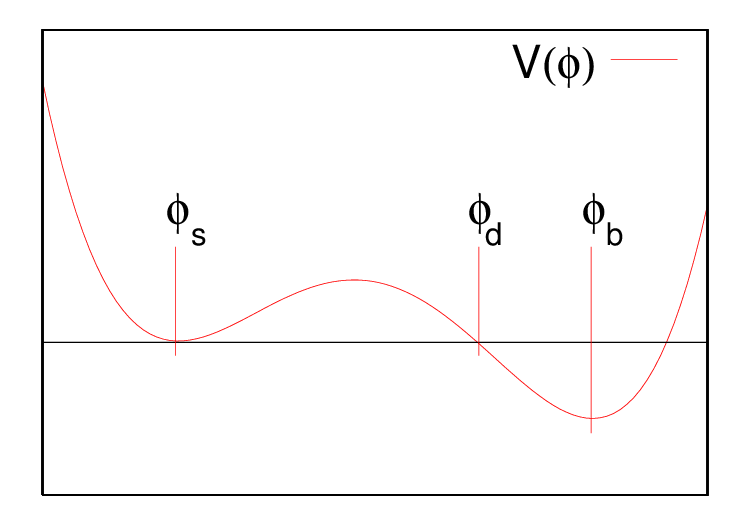}
\end{center}
\caption{
\label{fig:pot_example}
\small Example for a potential with a metastable minimum. The phase transition proceeds from the symmetric phase $\phi_s$ to the broken phase $\phi_b$. }
\end{figure}

In a tunneling problem the effective potential has at least two local minima that constitute the different phases the physical system can reside in. In the following we call these two phases the symmetric (before the phase transition) and broken (after the phase transition) phases, motivated by the electroweak phase transition, see Fig~\ref{fig:pot_example}.
In the semi-classical WKB approximation, the tunnel probability per volume and time is suppressed by the Euclidean action of the so-called tunneling bounce $\bar \phi$ 
\be
\label{eq:tunnel_prop}
P \sim A \, e^{-S(\bar \phi)}\, ,
\ee
derived from the effective action expanded in gradients
\be
\label{eq:eff_action}
S \simeq \int d^4x \,  \frac12 \partial^\mu \phi \partial_\mu \phi + V(\phi) \, ,
\ee
where $V(\phi)$ denotes the corresponding effective potential that eventually depends on the temperature.

The coefficient $A$ in (\ref{eq:tunnel_prop}) is for dimensional reasons of electroweak scale, $A \sim T^4$. The phase transition happens when the probability to nucleate a bubble of broken phase is of order unity in a Hubble volume and time leading to the condition
\be
S \simeq \log \frac{A}{H^4} \simeq 140 \, .
\ee

The bounce $\bar \phi$ is at zero temperature a $O(4)$-symmetric solution to the Euclidean equations of motion while at finite temperature it is $O(3)$ symmetric and periodic in imaginary time. The equations of motion then read
\be
\frac{d^2 \bar \phi}{d\tau^2} 
+ (d-1) \frac{d \bar \phi}{\tau \, d\tau} 
= - \frac{dV}{d\bar \phi} \, ,
\ee
with $d=4$ ($d=3$) for tunneling at zero (finite) temperature. The boundary conditions are such that $\bar\phi$ initially rests close to the broken phase and asymptotically approaches the symmetric phase at late 'time'
\be
\bar \phi(0) \simeq \phi_b \, , \quad  
\bar \phi^\prime(0)=0 \, , \quad
\bar \phi(\infty)=\phi_s \, . \quad 
\ee

In the limit of weak phase transitions, the thin-wall approximation applies~\cite{Coleman:1977py}. In this case the field $\bar \phi$ rests for a rather long time $\tau_R$ close to the broken phase and then quickly changes to the symmetric phase. In this case, the tunnel action can be reexpressed in terms of the wall tension
\be
\sigma = \int d\phi \sqrt{2 V(\phi)} \, ,
\ee
and the potential difference $\Delta V \equiv V(\phi_b) - V(\phi_s)$ as 
\bea
S &=& \frac{27 \pi^2 \sigma^4}{2 \Delta V^3} 
\quad (d=4), \nn \\
S &=& \frac{16 \pi \sigma^3}{3 T \Delta V^2} 
\quad (d=3). 
\eea
Otherwise, for one field and quite arbitrary conditions, the tunneling action can easily be obtained numerically using the shooting-algorithms~\cite{Coleman:1977py}. For several scalar fields, more involved methods have to be used \cite{Cline:1999wi, Konstandin:2006nd}. 

Recently, the gauge-independence of above approach was questioned~\cite{Patel:2011th, Wainwright:2011qy, Wainwright:2012zn} but an explicit calculation in a Abelian toy model shows that the dependence on the gauge choice is actually quite small~\cite{Garny:2012cg}.
This is also supported by the fact that the semi-classical approach agrees reasonably well with non-perturbative methods on the lattice~\cite{Moore:2000jw}. The main corrections to the procedure above seem to come from higher order contributions to the kinetic term and the effective potential in the effective action (\ref{eq:eff_action}). 

\section{Wall velocity and wall thickness}

Several parameters of the phase transition enter the produced baryon asymmetry quantitatively. Namely the critical vev $\phi_c$, the critical temperature $T_c$, the wall thickness $\ell_w$ and the wall velocity $v_w$. 

The most important one is hereby the ratio $\phi_c/T_c$ that determines the sphaleron washout and also the reflection of the particles by the Higgs wall that in turn leads to the CP violation in the particle densities. Fortunately, in most models with viable electroweak baryogenesis this quantities can be rather easily obtained using the semi-classical methods of appendix~\ref{sec:app_tunnel}.

Another important input is the wall thickness $\ell_w$. The gradient expansion can only be applied for thick walls, $\ell_w \, T \gg 1$, and the final baryon asymmetry is in the one-flavor case roughly inversely proportional to the wall thickness. For not too large wall velocities, the shape of the Higgs bubble profile does not change much during the expansion~\cite{Espinosa:2010hh}. The wall thickness can then be determined from the wall thickness of the nucleated bubbles in the semi-classical tunneling analysis. 

Finally, the wall velocity $v_w$ enters the analysis. Viable baryogenesis requires that the wall velocity is small enough to allow for particle diffusion in front on the wall $v_w < 1/\sqrt{3}$. For wall velocities smaller than that the produced baryon asymmetry is rather insensitive to the wall velocity as already discussed in appendix~\ref{sec:Gammaws}. This results from the fact that the CP violation accumulated in front of the wall is proportional to the wall velocity. At the same time, the phase transition proceeds longer and the sphaleron process can act longer on the CP-violating particle densities and convert them into a baryon asymmetry. In this regime the final baryon asymmetry depends only weakly the wall velocity. However, for very slow walls, the sphaleron process becomes saturated and the final asymmetry scales linearly with the wall velocity. Due to the smallness of the sphaleron rate, this typically happens for velocities of order $v_w \lesssim 10^{-3}$.

So the pivotal question is if the wall velocity is in the regime $10^{-3} \ll v_w < 1/\sqrt{3}$ where above approximations are reasonable and the final asymmetry is insensitive to the wall velocity. To answer this question in a specific model requires to perform an out-of-equilibrium analysis that so far was only performed in the Standard Model~\cite{Moore:1995ua, Moore:1995si} and the MSSM~\cite{John:2000zq}. In both cases, the wall velocity turned out to be in the desired ballpark. For other models, the wall velocity is still unknown. A simple way of estimating the wall velocity is to model friction in a phenomenological approach and to extrapolate the results from the SM and the MSSM~\cite{Megevand:2009ut, Megevand:2009gh, Espinosa:2010hh, Huber:2011aa, Huber:2013kj}.

\section{Electroweak phase transition in the SM\label{sec:PT_in_SM}}

In this section we review the perturbative analysis of the electroweak phase transition in the Standard Model. We follow the work~\cite{Anderson:1991zb} but present a simplified analysis.

At tree level, the effective potential of the Higgs field reads
\be
V^0 = \frac{\lambda}{4} \left( \phi^2 - v^2 \right)^2 \, ,
\ee
and at one loop order the thermal corrections to the free energy are
\be
\Delta V^1 = \mp \frac{T^4}{2 \pi^2} \sum_i \int \, dx \, x^2 \log 
\lp 1 \pm \exp( - \sqrt{x^2 + m_i^2 \beta^2 } \rp \, ,
\ee
where $\pm$ stand for fermions/bosons respectively, $T$ denotes the temperature, $\beta$ the inverse temperature and $m_i$ the different particle masses. As long as the masses do not exceed the temperature, this can be expanded as
\bea
\Delta V^1_{fermions} &=& 
\frac1{48} m^2 T^2 + O(m^4)\, , \nn \\ 
\Delta V^1_{bosons} &=&  
\frac1{24} m^2 T^2 - \frac{1}{12 \pi} m^3 T + O(m^4) \, .
\eea
Of special importance are hereby the cubic terms contributed by the bosons. 
If the mass of a bosonic field is only generated by the coupling to the Higgs vev (as is the case for the weak gauge bosons in the SM), this gives in turn rise to a term of the form $\phi^3 T$ in the effective potential. This term is essential to generate a potential barrier between the symmetric and the broken phase.
Consider a potential of the form
\be
\label{eq:Vpoly}
V = \mu^2(T) \phi^2 - E \, T \, \phi^3 + \frac{\lambda}{4} \phi^4 \, .
\ee
At some temperature $T_c$, this polynomial potential has two degenerate minima at $\phi=0$ and $\phi=\phi_c > 0$ and is of the form
\be
V = \frac{\lambda}{4} \phi^2 (\phi-\phi_c)^2 \, .
\ee
Comparison with (\ref{eq:Vpoly}) then shows that 
\be
\mu^2 (T_c) = \frac14 \lambda \phi_c^2\, , 
\quad E \, T_c = \frac12 \lambda \phi_c \, .
\ee
This immediately implies 
\be
\label{eq:PTstrength}
\phi_c / T_c = 2 E / \lambda 
\ee
and larger Higgs masses lead to weaker phase transitions.

In the Standard Model, the cubic coefficient arises only from the (transverse) electroweak gauge bosons~\cite{Carrington:1991hz, Dine:1992vs, Dine:1992wr}, $E \sim 10^{-2}$. Accordingly, a phase transition strong enough for electroweak baryogenesis is only possible for Higgs masses below $40$ GeV~\cite{Kajantie:1995kf} in light of the constraint (\ref{eq:washout}). Besides, for Higgs masses $m_h \gtrsim 70$ GeV the perturbative analysis breaks down and a cross-over replaces the phase transition. 

In the MSSM, additional contributions to the cubic term come from the right-handed stops in case their mass is below the top mass. This can make the phase transition strong enough for baryogenesis for some parts of the parameter space, even when a Higgs of mass $m_h \simeq 125$ GeV is assumed (see sec.~\ref{sec:MSSM}). 

In general, if the potential barrier arises from a thermal cubic contribution, the relation (\ref{eq:PTstrength}) in combination with a Higgs mass of $m_h \simeq 125$ GeV implies that a strong first-order phase transition requires at least $E \gtrsim 0.1$. So a moderate number of light bosons that couple strongly to the Higgs is in this case essential. Yet, in many models the strength of the phase transition does not rely on the thermal cubic contributions. The prime example for this are models with an extended Higgs sector. When several scalar fields acquire a vev at electroweak scales, potential barriers can arise even in the tree level scalar potential (see sec.~\ref{sec:LC2}).

\bibliography{biblio}
\bibliographystyle{JHEP}

\end{document}